\documentclass[12pt]{article}
\usepackage{pstricks}
\usepackage{color}
\usepackage{amssymb,amsmath,bm,bbm}
\usepackage{epsf}
\usepackage{epsfig}
\usepackage{afterpage}
\usepackage{longtable}
\usepackage{cite}
\usepackage{latexsym, mathrsfs}
\usepackage{graphics}
\usepackage{url}
\usepackage{paralist}
\usepackage{bbold}

\newcommand{\be}{\begin{equation}}
\newcommand{\ee}{\end{equation}}
\newcommand{\bea}{\begin{eqnarray}}
\newcommand{\eea}{\end{eqnarray}}
\newcommand{\nn}{\nonumber}
\newcommand{\dd}{\displaystyle}

\newcommand{\ket}[1]{\left| #1 \right\rangle}

\newcommand{\epsK}{\varepsilon_K}
\def\epe{\varepsilon'/\varepsilon}
\newcommand{\gev}{\rm GeV}
\newcommand{\mev}{\rm MeV}

\def\kpn{K^+\rightarrow\pi^+\nu\bar\nu}
\def\klpn{K_{L}\rightarrow\pi^0\nu\bar\nu}
\begin{document}
\begin{flushright}
{AJB-21-4}\\
    {BARI-TH/21-729}
\end{flushright}

\medskip

\begin{center}
{\LARGE\bf
\boldmath{ The charm of 331  }}
\\[0.8 cm]
{\bf Andrzej~J.~Buras$^a$, Pietro Colangelo$^{b}$, Fulvia~De~Fazio$^{b}$ and
Francesco~Loparco$^{b,c}$
 \\[0.5 cm]}
{\small
$^a$TUM Institute for Advanced Study, Lichtenbergstr. 2a, D-85747 Garching, Germany\\
$^b$Istituto Nazionale di Fisica Nucleare, Sezione di Bari, Via Orabona 4,
I-70126 Bari, Italy\\
$^c$Dipartimento Interateneo di Fisica "Michelangelo Merlin", Universit\`a degli Studi di Bari, via Orabona 4, 70126 Bari, Italy}
\end{center}

\vskip0.41cm


\abstract{%
We perform a detailed analysis of flavour changing neutral current processes
in the charm sector in the context of 331 models.
 As pointed out recently, 
 in the case of $Z^\prime$ contributions in these models 
there are no new free parameters beyond those already present  in the
$B_{d,s}$ and $K$ meson systems   analyzed 
in the past. As a result, definite ranges for new Physics (NP) effects
in various charm observables could be obtained. While generally NP
effects turn out to be small, in a number of observables they are much larger
than the tiny effects predicted within the Standard Model. In particular we find that the branching ratio of the mode $D^0 \to \mu^+ \mu^-$,  despite remaining tiny, can be enhanced by 6 orders of magnitude with respect to the SM. We work out correlations between this mode and rare $B_{d,s}$ and $K$ decays. 
We also discuss neutral charm meson oscillations and CP violation in the charm system. In particular, we point out that 
  331 models provide new weak phases that are a necessary condition to have non-vanishing CP asymmetries.
In the case of  $\Delta A_{CP}$,   the difference between the CP asymmetries in $D^0 \to K^+ K^-$ 
and $D^0 \to \pi^+ \pi^-$, we find that agreement with experiment  can be obtained provided that two conditions are verified: the phases  in the ranges predicted in 331 models and large hadronic matrix elements.
\noindent}


\section{Introduction}
The role of charm in particle physics can hardly be overemphasized\cite{Wilkinson:2021tby}. 
Not only historically  the discovery of $ J/\psi$,  and hence of the charm quark, was so important to be ascribed the name of {\it November revolution}, but even in much more recent times, the observation of charmed  states with properties not expected in the quark model requires the development of a new  hadron spectroscopy, to incorporate all hints of exotic states in a  predictive framework. 

However, the peculiar role of charm brings in also difficulties from the theory point of view. The first reason lies in the fact that the charm quark  is not as heavy as the beauty quark to allow a completely reliable use of the heavy quark techniques efficiently developed for beauty hadrons. On the other hand, it cannot either be considered a light quark, so that the theoretical predictions  are very often affected by uncertainties difficult to control because of long distance dynamics.

Among the observables in the charm sector showing this kind of problems there  are the neutral meson mixing, i.e. $D^0-{\overline D}^0$ oscillation, and CP violation, as well as rare charm decays  governed by penguin diagrams \cite{Lenz:2020awd}.
Two facts make the Standard Model (SM) treatment of these processes rather challenging.
The first one is that  loop-induced processes in the charm case get contributions from internal down-type quarks. Since the loop functions depend on the ratios $x_i=m_i^2/M_W^2$, the smallness of these ratios for $i=d,\,s,\,b$ makes them all very close to zero and therefore almost equal,  resulting in almost perfect   GIM cancellation.
 Another point is that the CKM elements involving charm have tiny imaginary parts,  which is a challenge for CP violation studies if the SM was the whole story. 
 Yet, we know that the SM is not the whole story. Hence, the strong suppression
   of some processes within  SM, namely lepton flavour violating decays and the
   electric dipole moments, can even be an advantage in the search for new
   physics (NP)  because of the small SM background.
In the case of  rare  flavour changing neutral current (FCNC) processes induced by the $c \to u$ transition the effectiveness of the GIM mechanism results in so tiny   branching ratios in the SM that one can consider them as {\it null tests}: their observation  with a less severe degree of suppression with respect to SM  would be a signal of NP.

Deviations from the SM predictions in the flavour sector have been observed in several cases: they  are usually referred to as {\it flavour anomalies}. Such tensions  deal with  loop-induced processes,  as in the ratios $R_{K^{(*)}}$ and in the so-called $P_5^\prime$ distribution  in $b \to s \ell^+ \ell^-$ processes, and very recently in  $(g-2)_\mu$. They also  affect tree-level charged current processes,  as in the ratios $R(D^{(*)})$ in  semileptonic $B$ decay and in the analogous ratios involving corresponding decays of other beauty hadrons \cite{Altmannshofer:2021qrr,Alguero:2021anc}.
In view of this, it is natural to look for deviations in other processes, and the purpose of this paper is  to
 explore possible deviations from the SM expectations in the charm sector. A SM extension that could be promising from this point of view is the 331 model, whose main features  we briefly  describe in the next Section. The reason is the presence in this model of a new neutral gauge boson $Z^\prime$  that can mediate tree-level FCNC. A few new parameters are introduced, affecting the interaction Lagrangian of SM fermions with $Z^\prime$. As observed in \cite{Colangelo:2021myn},  it is remarkable that the parameters  governing the flavour violating  coupling $cuZ^\prime$, which enter in FCNC charm transitions, up to CKM factors are the same  parameters  describing  the $sdZ^\prime$  and $bsZ^\prime$ couplings. This is a rather powerful feature: it allows us to exploit the experimental constraints in the kaon and in the beauty systems to place bounds on  rare charm transitions and to establish  correlations among observables in the three sectors. 

The plan of the paper is the following. In Section \ref{331} we present the basic features of the 331 model and of its variants.  Section \ref{charm} is devoted to   $D^0-{\overline D}^0$ mixing, CP violation  and  rare charm decays in the SM and in 331 models. The numerical results and  the correlations with kaon and $B$ observables are presented in Section \ref{numerics}. In the last Section we summarize our findings. 

\section{331 model}\label{331}
The extension of the SM gauge group to $SU(3)_c \times SU(3)_L \times U(1)_X$,  proposed in \cite{Pisano:1991ee,Frampton:1992wt}, defines a class of models generically referred to as  
 331  models. In these models two spontaneous symmetry breakings occur: first to the SM group 
  $SU(3)_c \times SU(2)_L \times U(1)_Y$ and then   to $SU(3)_c \times U(1)_Q$. 
 The larger group leads to the introduction of  5 additional gauge bosons, as well as new fermions. Indeed, with the enlargement $SU(2)_L \to SU(3)_L$ the  left-handed SM fermions are components of triplets or antitriplets rather than doublets. As a consequence, a third partner is usually needed,  in general a new heavy fermion.

 A very appealing  feature of 331 models is that 
 the requirement of anomaly cancelation together with that of asymptotic freedom of QCD 
constrains  the number of generations to be equal to the number of colours, 
making the 331 able to predict  the existence of just three generations of
ordinary quarks\footnote{This is not possible within the SM unless other
  constraints like the width of Higgs are taken into account.}.
Moreover, quark generations should transform differently under the action of $SU(3)_L$. In particular, two quark generations  should transform as triplets, one as an antitriplet. 
The common choice is that  the latter is
 the third generation.  The different role of the third generation with respect to the others could  possibly be at the origin of the large top quark  mass. 
 
The relation
    \be
   Q=T_3+\beta T_8+X
   \ee
   between the electric charge $Q$ and 
  $T_3$ and $T_8$, two of the $SU(3)$ generators, and $X$, the generator of $U(1)_X$, introduces a parameter  $\beta$. This  is a key parameter that defines the specific variant of the model. 
  Among the new gauge bosons, four of them $Y^{Q_Y^\pm}$ and $V^{Q_V^\pm}$ have properties that
 depend on the value of $\beta$: 
they  have integer charges when  $\beta$  is multiple of $\displaystyle{\frac{1}{\sqrt{3}}} $ and of $\sqrt{3}$. 
On the other hand, 
in all the model variants a new neutral gauge boson $Z^\prime$ is present. A very appealing feature of the 331 models  is that $Z^\prime$   mediates tree-level flavour changing neutral currents  in the quark sector, while its couplings to leptons are  diagonal and universal. 
An extended Higgs sector  also comprises  three $SU(3)_L$ triplets and one sextet.
Finally, a relation exists between 
the $U(1)_X$ gauge coupling $g_X$ and the $SU(3)_L$ coupling $g$:
\be
{g_X^2 \over g^2}={6 \sin^2 \theta_W \over 1-(1+\beta^2) \sin^2 \theta_W}\,\,,
\ee
where $\theta_W$ is the  Weinberg angle.

In analogy to the SM, quark mass eigenstates are obtained through rotation of flavour eigenstates by means of two unitary matrices $U_L$ (for up-type quarks) and $V_L$ (for down-type ones) that satisfy the relation $V_{CKM}=U_L^\dagger V_L$, as in the SM case. However, while in  the SM $V_{CKM}$ appears only in charged current interactions and $U_L$ and $V_L$ never appear individually, in the 331 models it is possible to get rid  of only one matrix, $U_L$ or $V_L$, by expressing it   in terms of $V_{CKM}$ and the other one; the remaining rotation matrix enters in the $Z^\prime$ couplings to quarks. Choosing $V_L$ as the surviving rotation matrix, one can   parametrize it as \cite{Buras:2012dp}
 \begin{equation}
V_L=\left(\begin{array}{ccc}
{\tilde c}_{12}{\tilde c}_{13} & {\tilde s}_{12}{\tilde c}_{23} e^{i \delta_3}-{\tilde c}_{12} {\tilde s}_{13} {\tilde s}_{23}e^{i(\delta_1
-\delta_2)} & {\tilde c}_{12}{\tilde c}_{23} {\tilde s}_{13} e^{i \delta_1}+ {\tilde s}_{12} {\tilde s}_{23}e^{i(\delta_2+\delta_3)} \\
-{\tilde c}_{13} {\tilde s}_{12}e^{-i\delta_3} & {\tilde c}_{12}{\tilde c}_{23} + {\tilde s}_{12}
 {\tilde s}_{13} {\tilde s}_{23}e^{i(\delta_1-\delta_2-\delta_3)} & - {\tilde s}_{12} {\tilde s}_{13}{\tilde c}_{23}e^{i(\delta_1 -\delta_3)}
-{\tilde c}_{12} {\tilde s}_{23} e^{i \delta_2} \\
- {\tilde s}_{13}e^{-i\delta_1} & -{\tilde c}_{13} {\tilde s}_{23}e^{-i\delta_2} & {\tilde c}_{13}{\tilde c}_{23}
\end{array}\right) \,\,.\label{VL-param}
\end{equation}
In the 331 Lagrangian density the term describing the $Z^\prime$ interaction with ordinary fermions  reads:
\bea\label{ZprimeFR}
i\,L_{int}^{Z^\prime} &=&  i{g  Z^{\prime \mu}\over 2 \sqrt{3} c_W \sqrt{1-(1+\beta^2) s_W^2}} \nn\\
 && \Bigg\{ \sum_{\ell=e,\mu,\tau}\Big\{\left[1-(1+\sqrt{3} \beta) s_W^2 \right] \left({\bar \nu}_{\ell \, L} \gamma_\mu \nu_{\ell \, L}+ {\bar \ell}_L \gamma_\mu \ell_L \right)- 2 \sqrt{3} \beta s_W^2 {\bar \ell}_R \gamma_\mu \ell_R 
\Big\}  \nn \\
&+&
\sum_{i,j=1,2,3} \Big\{ \big[-1+(1+{\beta \over \sqrt{3}})s_W^2\big]({\bar q}_{uL})_i \gamma_\mu (q_{uL})_j \delta_{ij}+ 2c_W^2 ({\bar q}_{uL})_i \gamma_\mu (q_{uL})_j u_{3i}^* u_{3j} \nn \\
&+& \big[-1+(1+{\beta \over \sqrt{3}})s_W^2\big]({\bar q}_{dL})_i \gamma_\mu (q_{dL})_j \delta_{ij}+ 2c_W^2 ({\bar q}_{dL})_i \gamma_\mu (q_{dL})_j v_{3i}^* v_{3j} \nn \\
&+&{4 \over \sqrt{3}}\beta s_W^2({\bar q}_{uR})_i \gamma_\mu (q_{uR})_j \delta_{ij}
-{2 \over \sqrt{3}}\beta s_W^2({\bar q}_{dR})_i \gamma_\mu (q_{dR})_j \delta_{ij}\Big\}
\,\,,
\eea
where $s_W\equiv \sin \theta_W,$ $c_W\equiv \cos \theta_W$,  $q_u \, (q_d)$ denotes an up (down)-type quark ($i,\,j$ are generation indices), and $v_{ij}$,  $u_{ij}$ are the elements of the  $V_L$ and $U_L$ matrices, respectively.

Using the parametrization in Eq.~(\ref{VL-param}) and the Feynman rules for $Z^\prime$ couplings to quarks from Eq.~(\ref{ZprimeFR}) \cite{Buras:2012dp,Buras:2013dea},
one finds   that the $B_d$ system involves only the parameters ${\tilde s}_{13}$ and $\delta_1$, while the $B_s$ system involves only
${\tilde s}_{23}$ and $\delta_2$. 
Kaon physics depends on  ${\tilde s}_{13}$, ${\tilde s}_{23}$ and $\delta_2 - \delta_1$, so that stringent correlations between observables in $B_{d,s}$ sectors and in the kaon sector can be established.

Exploiting this  feature,   correlations among several quark flavour observables in $B_d$, $B_s$ and $K$ decays have been studied in \cite{Buras:2012dp,Buras:2013dea,Buras:2014yna,Buras:2015kwd,Buras:2016dxz}.
Moreover,  the relation 
\be
U_L=V_L \cdot V_{CKM}^\dagger
\label{UL}
\ee
produces  additional correlations between  $B_d,\,B_s,\,K$ decays and the
  transitions induced by up-type quark FCNC decays mediated by $Z^\prime$, which 
 occur  in 331 models.
In particular, as it follows from Eq.~(\ref{ZprimeFR}),  correlations exist  between observables in the charm sector and  observables for $B_d$, $B_s$ and $K$. This observation has been exploited  to relate the $c \to u \nu \bar \nu$ processes, such as   $B_c \to B_u^{(*)} \nu {\bar \nu}$, to processes induced by $b \to s \nu {\bar \nu}$ and $s \to d \nu {\bar \nu}$  \cite{Colangelo:2021myn}. Here we focus on other important processes in the charm sector.

As shown in Ref.\cite{Buras:2013dea}, the variants of the model with $\beta=\pm\displaystyle\frac{2}{\sqrt{3}}$ and $\beta=\pm\displaystyle\frac{1}{\sqrt{3}}$, with the fermions in the third generation  transforming as antitriplets, satisfy a number of phenomenological constraints, such as those   from $\Delta F=2$ observables in the $B_d,\,B_s, \,K$ systems, as well as the electroweak precision observables,  provided that the mass of $Z^\prime$ is not smaller than about $1$ TeV. Among such variants, 
 the one with  $\beta=\displaystyle\frac{2}{\sqrt{3}}$  (denoted as M8 in \cite{Buras:2015kwd})  predicts relevant contributions to the ratio 
$\varepsilon^\prime/\varepsilon$ \cite{Buras:2015kwd}. 

Another remark concerns the  $Z-Z^\prime$ mixing  in this model  \cite{Buras:2014yna}.
The mixing  can be neglected in the case of $\Delta F=2$ transitions and in decays like $B_d \to K^* \mu^+ \mu^-$, where they are suppressed by the small vectorial coupling of $Z$ to charged leptons.
 However,
the contributions of tree-level $Z$ exchanges to decays sensitive to axial-vector couplings,
namely  $B_{s,d} \to \mu^+ \mu^-$,   $B_d \to K \mu^+ \mu^-$ and those with neutrinos in the final state, cannot be neglected, in general.  The mixing angle can be written as \cite{Buras:2014yna}:
\be\label{sxi}
\sin\xi=\frac{c_W^2}{3} \sqrt{f(\beta)}\left(3\beta \frac{s_W^2}{c_W^2}+\sqrt{3}a\right)\left[\frac{M_Z^2}{M_{Z^\prime}^2}\right]
\equiv B(\beta,a) \left[\frac{M_Z^2}{M_{Z^\prime}^2}\right],
\ee
where 
\be\label{central}
f(\beta)=\frac{1}{1-(1+\beta^2)s_W^2} > 0\,. 
\ee
The parameter $a$ introduced in Eq.~\eqref{sxi} is defined in terms of 
the 
vacuum expectation values of two  Higgs triplets $\rho$ and $\eta$, as follows:
\be\label{ratiov}
-1 < a=\frac{v_-^2}{v_+^2}< 1 \ee
\be
v_+^2=v_\eta^2+v_\rho^2,  \qquad  v_-^2=v_\eta^2-v_\rho^2\,.
\ee
With the same notation of  the  two Higgs doublet models,   $a$  is expressed in terms of the parameter $\tan\bar\beta$   as \footnote{The  overline  distinguishes  $\bar \beta$  from the 
parameter $\beta$ characterizing the 331 models.}:
\be\label{basica}
a=\frac{1-\tan^2\bar\beta}{1+\tan^2\bar\beta}, \qquad \tan\bar\beta=\frac{v_\rho}{v_\eta}.
\ee
A detailed analysis of the impact of the $Z-Z^\prime$ mixing  in several 331 models  distinguished by different values of   $\beta$, including the
  constraints from electroweak precision observables, is presented in  \cite{Buras:2014yna}.  Here, we analyze the impact on the charm sector, in particular on $D^0-{\overline D}^0$ mixing, on  CP violation in $D^0$ system and on rare $D^0$ 
  transitions \footnote{The mass difference in the neutral D  system has been  studied in  331 models with $\beta=\pm 1/\sqrt{3}$  considering the contribution to FCNC of the mixing between ordinary and exotic quarks   \cite{Cabarcas:2009vb}. The effects of new scalars have been investigated in \cite{Machado:2013jca}, while
 FCNC processes in 331 models with right-handed neutrinos have been considered in \cite{Cogollo:2012ek}. }.

\section{Charm observables in the SM and in 331 model}\label{charm}
\subsection{$D^0-{\overline D}^0$ mixing and $x_D$}
Neutral charmed mesons provide the only system where flavour oscillation of up-type quarks can occur. 
The time evolution of the $D^0-{\overline D}^0$ system is governed by a Schr\"odinger  equation:
\be
\label{mixing}
i\frac{\partial}{\partial t} \begin{pmatrix}D^0 \\ {\overline D}^0 \end{pmatrix} ={\cal M} \begin{pmatrix}D^0 \\ {\overline D}^0 \end{pmatrix}=
\addtolength{\arraycolsep}{3pt}
\left(\begin{array}{cc} M_{11}^D -\frac{i}{2}\Gamma_{11}^D &
 M_{12}^D -\frac{i}{2}\Gamma_{12}^D \\
 {M_{12}^D}^* -\frac{i}{2}{\Gamma_{12}^D}^* &
 M_{11}^D -\frac{i}{2}\Gamma_{11}^D 
\end{array}\right)\addtolength{\arraycolsep}{-3pt}
\begin{pmatrix}D^0 \\ {\overline D}^0 \end{pmatrix}\,.
\ee
Due to off-diagonal terms in 
${\cal M}$, mass eigenstates do not coincide with the flavour eigenstates $D^0,\,{\overline D}^0$, but are given by 
\bea
| D_1 \rangle &=& \frac{1}{\sqrt{|p|^2+|q|^2}} \left( p | D^0 \rangle + q | {\overline D}^0 \rangle \right)\,,\\
| D_2 \rangle &=& \frac{1}{\sqrt{|p|^2+|q|^2}} \left( p | D^0 \rangle - q | {\overline D}^0 \rangle \right) \,,
\eea
with
\be\label{qoverp}
\frac{q}{p} \equiv \sqrt{\frac{{{M_{12}^D}^*} - \frac{i}{2}{{\Gamma_{12}^D}^*}}{M_{12}^D - \frac{i}{2}\Gamma_{12}^D}}=\left|\frac{q}{p} \right| e^{\,i \phi} \,.
\ee
CP parity transforms $D^0,\,{\overline D}^0$ into each other modulo a phase that can be chosen in such a way that 
\be\label{CPD0}
CP |D^0\rangle = + |{\overline D}^0\rangle\,.
\ee
Mass and width differences are usually normalized to the average width of the two mass eigenstates:
\be
x_D =\frac{\Delta M_D}{\overline \Gamma} \, , 
\qquad y_D= \frac{\Delta \Gamma_D}{2\overline \Gamma} \, , \qquad 
\overline \Gamma = \frac{1}{2} (\Gamma_1 + \Gamma_2) \, , \label{xDyD}
\ee
with 
\bea
\Delta M_D&=&M_1-M_2=2 \,{\rm Re} \left[\frac{q}{p}{\cal M}_{12} \right] \\
\Delta \Gamma_D&=&\Gamma_1-\Gamma_2=-4 \,{\rm Im} \left[\frac{q}{p}{\cal M}_{12} \right] .
\eea
%
%
 In the 2021 online report the HFLAV group  quoted the  summary of the charm-mixing data  obtained  from global fits in which CP violation is allowed \cite{Amhis:2019ckw}:
\bea
x_D &=&\big(3.7\pm1.2 \big) \times 10^{-3} \, , 
\qquad y_D= \big(6.8^{+0.6}_{-0.7} \big) \times 10^{-3} \, ,
 \label{xDyDexp}\\
\Big | \frac{q}{p}\Big | &=& 0.951^{+0.053}_{-0.042} \, , \hskip 2.3 cm  
\phi=(-5.3^{+4.9}_{-4.5})^\circ \, .
\label{xDyDexp1}
\eea
%
  New measurements  have been provided by the LHCb collaboration
 allowing for CP violation in mixing and in the interference between mixing and  decay. The best fit point is \cite{Aaij:2021aam}:
\bea
x_D &=&\big(3.98^{+0.56}_{-0.54} \big) \times 10^{-3} \, , 
\qquad y_D= \big(4.6^{+1.5}_{-1.4} \big) \times 10^{-3} \, ,
 \label{xDyDexpnew1}\\
\Big | \frac{q}{p}\Big | &=& 0.996\pm0.052 \, , \hskip 1.8 cm  
\phi=(0.056^{+0.047}_{-0.051}) \, .
\label{xDyDexp1new1}
\eea

It appears then that in the $D$ system $x_D\sim y_D$ or $\Gamma_{12}\sim M_{12}$, 
whereas in the $B$ system $|\Gamma_{12}/ M_{12}|\ll 1$. For  future prospects see \cite{Bhardwaj:2019vep,Cerri:2018ypt}. 
In the limit of (approximate) CP~symmetry $x_D$, $y_D > 0$ implies that the CP~{\it even} state ($D_1$) is 
slightly heavier and shorter lived than the CP~{\it odd} one ($D_2$) (unlike for neutral kaons).

 In the SM the $D^0-{\overline D}^0$ mixing originates in the box diagrams with $W$ exchange and internal down-type quarks, as opposed to  $B_{d,s}^0-{\overline B}_{d,s}^0$ and  $K^0-{\overline K}^0$ mixings where the up-type quarks are exchanged,   in particular the top-quark. As a result, the mass differences $\Delta M_{s,d}$ predicted by the SM turn out to be in the ballpark of their experimental   values.   This also holds for $\Delta M_K$, although in this
  case sizable long-distance contributions are present which are still not
  fully under control \cite{Cerri:2018ypt}.

  As in the case of $B_{d,s}^0-{\overline B}_{d,s}^0$ and  $K^0-{\overline K}^0$ mixings, the short-distance contribution to  $D^0-{\overline D}^0$ mixing in the SM is governed by a single operator $Q_1=({\bar u}_L \gamma_\mu c_L)({\bar u}_L \gamma_\mu c_L)$,  so that this contribution can be easily evaluated.
  One finds  that, due to strong GIM suppression, $x_D$ and $y_D$ are by
  at least one order of magnitude below the experimental values in (\ref{xDyDexp}), so that either  long-distance or new physics contributions  are responsible for their measured values. 

Many predictions   for $x_D,\,y_D$ in the SM  have been worked out, which however are affected by large uncertainties, as discussed  in Section~3 of  \cite{Cerri:2018ypt}. As in the SM,
$M_{12}$ corresponds to the dispersive part and $\Gamma_{12}$ to the absorptive part of the box diagrams, which are both strongly suppressed by the GIM mechanism.
The short-distance SM contributions to them are 
both real to a very good approximation, and this is also the case of the long-distance contributions. This implies that, unless NP contributions to $x_D$ and $y_D$ in a given model are in the ballpark of their values in  (\ref{xDyDexp}), the
best chance to search for NP in $D^0 - {\overline D}^0$ mixing is through CP violation, represented
by the departure of $|q/p|$ from unity and the non-vanishing phase $\phi$ in the measured range  in (\ref{xDyDexp1}).

If NP is taken into account, other operators beyond $Q_1$ can contribute.
The experimental values in (\ref{xDyDexp}) and (\ref{xDyDexp1}) provide  strong
constraints to the  extensions of the SM in which left-right, scalar
and tensor operators are present  \cite{Gabbiani:1996hi,Bona:2007vi,Isidori:2010kg,Charles:2013aka}. 
Indeed, such operators have
large hadronic matrix elements and strongly enhanced Wilson coefficients
through QCD renormalization group effects,

In  the 331 model only the  operator $Q_1$ is involved,  as in the SM,   but this time its Wilson coefficient  receives the contribution from  tree-level $Z^\prime$ exchange. However,  even such tree-level contributions
 are below the experimental values of $x_D$ and $y_D$ when the contraints from the other quark sectors are included.
As a matter of fact in the 331 model,  if all other contributions to the mixing are ignored,  
 at the scale $M_{Z^\prime}$  $x_D$ is given by:
\be
x_D^{331}(M_{Z^\prime})=\frac{F_D^2 m_{D^0} B_1}{2M_{Z^\prime}^2 \, \Gamma_D}\frac{2}{3}|\Delta_L^{uc}|^2 \,\,,
\label{xD331}
\ee
where $F_D$ is the $D^0$ decay constant and  $B_1$ parametrizes the deviation of the matrix element of the four quark operator $Q_1=({\bar u}_L \gamma_\mu c_L)({\bar u}_L \gamma_\mu c_L)$ from the vacuum saturation approximation. The coupling 
$\Delta_L^{uc}(Z^\prime)$ reads:
\bea
\Delta_L^{uc}(Z^\prime)&=&\frac{g \, c_W}{\sqrt{3} \, \sqrt{1-(1+\beta^2) s_W^2}}u_{31}^* u_{32}\,.
\label{Duc}
\eea
The elements $u_{ij}$ can be obtained using Eqs.~(\ref{VL-param}) and (\ref{UL}):
\bea
u_{31}&=& \sqrt{1 - {\tilde s}_{13}^2}  \sqrt{1 - {\tilde s}_{23}^2} V_{ub}^* - 
e^{-i \, \delta_1} {\tilde s}_{13} V_{ud}^*- 
 e^{-i \delta_2}\sqrt{1 - {\tilde s}_{13}^2}   {\tilde s}_{23} V_{us}^*
 \\
 u_{32}&=& \sqrt{1 - {\tilde s}_{13}^2}  \sqrt{1 - {\tilde s}_{23}^2} V_{cb}^* - 
e^{-i \, \delta_1} {\tilde s}_{13} V_{cd}^*- 
 e^{-i \delta_2}\sqrt{1 - {\tilde s}_{13}^2}   {\tilde s}_{23} V_{cs}^* \,\,.
 \eea
Hence,  $x_D^{331}$ depends on the four parameters ${\tilde s}_{13},\,\delta_1$ and 
 ${\tilde s}_{23},\, \delta_2$   which govern $B_d$ and $B_s$ decays, respectively, and which altogether  govern the $K$ decays.
 
 When evolved down to $m_c$, the  RG running gives \cite{Golowich:2007ka,Golowich:2009ii}:
 \be
 x_D^{331}(m_c)= r(m_c,\,M_{Z^\prime})\, x_D^{331}(M_{Z^\prime}) \label{xD331evol}
 \ee
 with
 \be
 r(m_c,\,M_{Z^\prime})=\left( \frac{\alpha_s(m_b)}{\alpha_s(m_c) } \right)^{6/25}\,\,
 \left( \frac{\alpha_s(m_t)}{\alpha_s(m_b) } \right)^{6/23} \,\,
 \left( \frac{\alpha_s(M_{Z^\prime})}{\alpha_s(m_t)} \right)^{2/7} \,\, .
 \ee
The
  evolution can be performed including NLO QCD corrections by means of the
  results in \cite{Buras:2000if}, but here we  restrict the analysis to the
  leading order.    The reason is that, as discussed in Section~\ref{numerics}, the  numerical value
  of $x_D^{331}$  resulting from \eqref{xD331},\eqref{xD331evol} is below the experimental measurement in
  (\ref{xDyDexp})  by about one order of magnitude when the constraints on ${\tilde s}_{13}$, $\delta_1$, 
 ${\tilde s}_{23}$ and $\delta_2$ are considered.

The small value of $x_D^{331}$  leads us  to conjecture that both SM  long-distance  and 331 contributions are responsible for the measured values of $x_D$ and $y_D$. The 331 term is pivotal in providing the complex phases producing  a deviation of $|q/p|$ from unity and a non-vanishing mixing phase $\phi$.
This is at the basis of the  analysis strategy in Section \ref{numerics}: 
   We use the measured values of $x_D$ and $y_D$  in Eq.~(\ref{xDyDexp}), interpreting them as resulting from the contributions of 
   both LD in the SM and of 331. Then, we predict  
   observables and correlations, as those  involving $|q/p|$ and $\phi$.

\subsection{CP asymmetries}
CP violation in the charm system can be tested in various ways. Here
we consider the time-dependent  CP asymmetries
\be\label{CPt}
A_{\rm CP}(f,t)\equiv\frac{\Gamma(D^0(t) \to f) -\Gamma({\overline D}^0(t) \to  f)}{\Gamma(D^0(t) \to f) + \Gamma({\overline D}^0(t) \to f)} \,\, , 
\ee
where $f$ is a CP eigenstate 
\be
{CP} |f\rangle = \eta_f |f\rangle \,,\qquad \eta_f=\pm 1\,.
\ee
Denoting
${\cal A}_f={\cal A}(D^0 \to f) $, ${\overline {\cal A}}_f={\cal A}({\overline D}^0 \to f) $
we define 
\be\label{eq:lf}
\lambda_f = \frac{q}{p}\frac{{\overline {\cal A}}_f}{{\cal A}_f}\, ,
\qquad \varphi_f = \frac{1}{2}{\rm Arg}\left(\lambda_f\right)\,.
\ee
The phase $\varphi_f$ is phase-convention independent
since it depends only on the relative phase between $q/p$ and ${\overline {\cal A}}_f/{\cal A}_f$:
\be
 \dd \frac{{\overline {\cal A}}_f}{{\cal A}_f}=\eta_f\left|\frac{{\overline {\cal A}}_f}{{\cal A}_f}\right|e^{-i2\xi_f} \, \, , \qquad \frac{q}{p} =\left|\frac{q}{p}\right|e^{i\phi}\, ,\qquad 
2\varphi_f=\phi-2\xi_f -{\rm Arg}(\eta_f)\,\, . \qquad 
\ee
\noindent 
%
where ${\rm Arg}(\eta_f)=0$ if $\eta_f=1$ and ${\rm Arg}(\eta_f)=\pi$ if $\eta_f=-1$.

In the {case of a non-negligible 
CP phase $\xi_f$ in the decay amplitude $ {\cal A}(D^0\to f)$, but  $| {\cal A}(D^0\to f)|= | {\cal A}(\bar D^0\to f)|$,} $\lambda_f$ simplifies
to
\be
\lambda_f = \eta_f \frac{q}{p} e^{-i2\xi_f},\qquad 
\left|\lambda_f\right|=\left|\frac{q}{p}\right|.
\ee
 In this case, as $x_D,y_D\ll 1$, it is appropriate to consider the CP~asymmetry in the limit of a small $t$  \cite{Bigi:2009df}:
\be\label{eq:smallt}
A_{\rm CP}(f,t)= -  \left[y_D \left(\left|\lambda_f\right|- 
\left|\frac{1}{\lambda_f}\right|\right)\cos2\varphi_f 
- x_D \left(\left|\lambda_f\right|+ \left|\frac{1}{\lambda_f}\right|\right)\sin2\varphi_f \right]\frac{t}{2\bar\tau_D}\,,
\ee
where $\bar\tau_D=1/\bar\Gamma$.

 On the other hand, when the phase $\xi_f$ is negligible, as happens in the SM and, as we shall see,  also  in the 331 models, one has
\be
\lambda_f = \eta_f \frac{q}{p} =
\eta_f\left|\frac{q}{p}\right|e^{i\phi}\,
\ee
and
\be
A_{\rm CP}(f,t)
\equiv S_{f} \frac{t}{ 2\overline\tau _D} \,,
\label{eq:GASYM}
\ee
with
\be
S_f\simeq - \eta_f  \left[ y_D\left(\left| \frac{q}{p}\right| -\left| \frac{p}{q}\right|   \right)\cos\phi -
x_D \left(\left| \frac{q}{p}\right| +\left| \frac{p}{q}\right|   \right) \sin\phi \right] \,.
\label{eq:Sf}
\ee
$S_f$ is analogous to $S_{\psi K_S}$ and $S_{\psi\phi}$ in the $B_d$ and $B_s$ system, respectively. However, in the $B$ system $y \ll x$ and $|q/p|\simeq1$, so that the above result simplifies considerably, leaving only the second term and allowing
the measurement of the phase $\phi$. 

Finally, we consider  the semileptonic asymmetry to  {\it wrong sign} leptons
\be\label{eq:ASL}
a_\text{SL}(D^0) \equiv  \frac{\Gamma (D^0(t) \to \ell^-\bar\nu X^+) - \Gamma (\bar D^0 \to \ell^+\nu X^-)}
{\Gamma (D^0(t) \to \ell^-\bar\nu X^+) + \Gamma (\bar D^0 \to \ell^+\nu X^-)}= 
\frac{|q|^4 - |p|^4}{|q|^4 + |p|^4}
\approx 2\left(\left|\frac{q}{p}\right|-1\right)
\ee
{with $X=K^{(*)}, \dots$, which represents CP~violation in mixing in $\Delta C=1$ transitions. The last
expression comes from the  condition  $\left| |{q}/{p}|-1 \right| \ll 1$.

\subsection{Correlations}
\label{sec:corr}

Two interesting correlations have  been derived in
\cite{Grossman:2009mn}, which read with our conventions  \cite{Bigi:2009df}:
\be\label{eq:varphi}
\sin^2\phi=\frac{x_D^2(1-|q/p|^2)^2}{x_D^2(1-|q/p|^2)^2+y_D^2(1+|q/p|^2)^2}.
\ee
In the limit $\big| |q/p|-1 \big| \ll 1$, $x_D\sim y_D$, the relation (\ref{eq:varphi}) simplifies to \cite{Grossman:2009mn}
\be
\sin\phi=\frac{x_D}{y_D}\left(1-\left|\frac{q}{p}\right|\right)\,.
\ee
Inserting this result in  (\ref{eq:Sf}) and (\ref{eq:ASL}) one finds for $\xi_f=0$:
\be
\label{eq:CORR1}
S_f=-\eta_f \frac{x_D^2+y_D^2}{y_D} a_\text{SL}(D^0)\,.
\ee
It should be emphasized that this relation is only valid if
  in our convention $\xi_f=0$ independently of $f$, that is no CP phase
in decay amplitudes. 
The experimental violation of the relation (\ref{eq:CORR1})  would then imply the occurrence  of CP violation in the decay
(direct CP violation)   \cite{Grossman:2009mn}. In the case of a significant phase $\xi_f$ one finds:
\be\label{eq:CORR2}
S_f=-\eta_f \left[ \cos2\xi_f \frac{x_D^2+y_D^2}{y_D}a_\text{SL}(D^0)
+2x_D \sin2\xi_f\right]\,.
\ee
As $\xi_f$ is phase convention dependent, the presence of direct CP violation
would be signalled by finding  $S_{f_1}-S_{f_2}\not=0$ for two different final states $f_1$ and $f_2$.

We shall analyze all these relations within 331 models in  Section~\ref{numerics}, but we want to have first a look at time-independent CP-asymmetries.

\subsection{$\Delta A_{CP}=A_{CP}(K^+K^-)-A_{CP}(\pi^+ \pi^-)$}
The first observation of CP violation in the decays of neutral  charm mesons was presented by the LHCb collaboration. For the difference
  between the CP asymmetries in $D^0\to K^+ K^-$ and $D^0\to \pi^+\pi^-$, 
\be
\Delta A_{CP}=A_{CP}(K^+K^-)-A_{CP}(\pi^+ \pi^-),
\ee
  the measurement is
 \cite{Aaij:2019kcg}
\be
\Delta A_{CP}=(-15.4\pm 2.9)\times 10^{-4}\,. \label{deltaACPexp}
\ee
With an excellent approximation this difference measures  direct CP violation. 
 Within the SM  this result has been explained invoking non perturbative effects enhancing the  penguin contributions  \cite{Golden:1989qx,Grossman:2006jg,Brod:2011re,Pirtskhalava:2011va,Bhattacharya:2012ah,Feldmann:2012js,Cheng:2012wr,Franco:2012ck,Cheng:2012xb,Cheng:2019ggx,Brod:2012ud}. Other studies envisage the possibility that 
 some NP could be responsible for this result
\cite{Isidori:2011qw,Giudice:2012qq,Altmannshofer:2012ur,Hiller:2012wf,Chala:2019fdb,Li:2019hho,Grossman:2019xcj,Dery:2019ysp}  (also supported by the analysis in  \cite{Khodjamirian:2017zdu}).   A discussion  of the LHCb data and of the future prospects
can be found in \cite{Lenz:2020awd}. Here we analyze the
asymmetry in 331 models.

The modes $D^0 \to K^+K^-$ and $D^0 \to \pi^+ \pi^-$ are single Cabibbo suppressed. 
These are  of the kind $D^0({\overline D}^0)  \to f$ with $f$ a  CP eigenstate that can be produced both in $D^0$ and ${\overline D}^0$ decays.
Denoting
${\cal A}_f={\cal A}(D^0 \to f) $, ${\overline {\cal A}}_f={\cal A}({\overline D}^0 \to f) $ and CP$\ket{f}=\eta_f \ket{f}$ ($\eta_f=\pm 1$), the 
CP asymmetry is defined as
\be
{A}_{CP}^f=\frac{\Gamma(D^0 \to f) -\Gamma({\overline D}^0 \to f)}{\Gamma(D^0 \to f) +\Gamma({\overline D}^0 \to f)} \,\,.
\ee
Note that in contrast to (\ref{CPt}) this asymmetry is time-independent.

When two amplitudes (at least) contribute to each decay, one can write \cite{Grossman:2006jg}
\be
{\cal A}_f={\cal A}_{f,1}+{\cal A}_{f,2}=|{\cal A}_{f,1}| \,e^{i \delta_{f,1}} e^{i \phi_{f,1}}+|{\cal A}_{f,2}| \,e^{i \delta_{f,2}} e^{i \phi_{f,2}}={\cal A}_{f,1}\left[1+r_f \, e^{i \delta_f} e^{i\ \phi_f} \right] \,\, , \label{Af}
\ee
where $r_f=|{\cal A}_{f,2}|/|{\cal A}_{f,1}|$ and $\delta_f=\delta_{f,2}-\delta_{f,1}$, $\phi_f=\phi_{f,2}-\phi_{f,1}$ are the difference between strong and weak phases, respectively.
If one of the amplitudes dominates, i.e. $r_f \ll 1$, one finds:
\be
{ A}_{CP}^f=-2 r_f \, \sin \delta_f \, \sin \phi_f \label{ACP} \,\,.
\ee
When NP is present the previous relations can be generalized by adding a new amplitude ${\cal A}_{f,NP}=|{\cal A}_{f,NP}|e^{i \, \delta_{NP}} e^{i \, \phi_{NP}}$ to (\ref{Af}):
\be
{\cal A}_f={\cal A}_{f,1}\left[1+r_f \, e^{i \delta_f} e^{i \phi_f}+{\tilde r}_f \, e^{i {\tilde \delta}_f} e^{i  {\tilde \phi}_f} \right]
 \ee
with ${\tilde r}_f=|{\cal A}_{f,NP}|/|{\cal A}_{f,1}|$ and ${\tilde \delta}_f=\delta_{NP}-\delta_{f,1}$, ${\tilde \phi}_f=\phi_{NP}-\phi_{f,1}$.
Consequently we have:
\be
{A}_{CP}^f=-2 r_f \, \sin \delta_f \, \sin \phi_f -2 {\tilde r}_f \, \sin{\tilde \delta}_f \, \sin {\tilde \phi}_f \label{ACPNPgen} \,\,.
\ee

In the SM both  $D^0 \to K^+K^-$ and $D^0 \to \pi^+ \pi^-$  have a tree-level contribution ${\cal A}_{SM}^{tree}$, proportional to $\lambda_s=V_{cs}^* V_{us}$ for  $f=K^+K^-$ and to $\lambda_d=V_{cd}^* V_{ud}$ for  $f=\pi^+ \pi^-$.  Penguin contributions are also present with internal down-type quark exchange and hence proportional to  $\lambda_D=V_{cD}^* V_{uD}$ with $D=d,s,b$. 
We denote such contributions as ${\cal A}_{SM}^{P,D}$. Exploiting the CKM unitarity relation $ \lambda_d+\lambda_s+\lambda_b=0$,  the general structure of the SM amplitude reads:
\bea
{\cal A}_K&=&{\cal A}_{K^+K^-}=\lambda_s({\cal A}_{SM,K}^{T}-{\cal A}_{SM,K}^{P,d})+\lambda_b({\cal A}_{SM,K}^{P,b}-{\cal A}_{SM,K}^{P,d}) \label{AKK}, \\
{\cal A}_\pi&=&{\cal A}_{\pi^+ \pi^-}=\lambda_d({\cal A}_{SM,\pi}^{T}-{\cal A}_{SM,\pi}^{P,s})+\lambda_b({\cal A}_{SM,\pi}^{P,b}-{\cal A}_{SM,\pi}^{P,s}) \label{Apipi} \,\,.
\eea

In 331 models  new contributions from tree-level $Z^\prime$ exchanges are present. Except for different hadronic matrix elements and the change of
the coupling $\Delta_L^{sd}(Z^\prime)$ to $\Delta_L^{uc}(Z^\prime)$ given  in Eq.~(\ref{Duc}), the structure of the $Z^\prime$ contribution to
the effective Hamiltonian at $M_{Z^\prime}$ is as in the case of the ratio
$\epe$  \cite{Buras:2014yna}.
 While flavour-changing $Z^\prime$ couplings involve only left-handed quarks, flavour conserving ones involve both left and right-handed ones. 
Flavour conserving $Z^\prime $ couplings to left-handed quarks are 
\bea
\Delta_L^{q_iq_i}(Z^\prime)&=&\frac{g }{2\sqrt{3} c_W}\sqrt{f(\beta)}\left\{ \left[-1+\left(1+\frac{\beta}{\sqrt{3}}\right)s_W^2\right]+2 c_W^2 u_{3i}^* u_{3i} \right\}
  \hskip .5cm  q_1=u,\,q_2=c \quad \quad \label{DLuu} 
   \\
   \Delta_L^{q_iq_i}(Z^\prime)&=&\frac{g }{2\sqrt{3} c_W}\sqrt{f(\beta)}\left\{ \left[-1+\left(1+\frac{\beta}{\sqrt{3}}\right)s_W^2\right]+2 c_W^2 v_{3i}^* v_{3i} \right\}\hskip .5 cm q_1=d,\,q_2=s
   \quad \quad \label{DLdd} \,\,,
   \eea
   with $f(\beta)$ given in (\ref{central}).
 Numerically, the second term in Eqs.~(\ref{DLuu})-(\ref{DLdd}) is  a factor of ${\cal O}(10^{-5})$ smaller than  the first one and  can be neglected,  so that we are left with the universal coupling
 \be
 \Delta_L^{qq}(Z^\prime)=\frac{g }{2\sqrt{3} c_W}\sqrt{f(\beta)} \left[-1+\left(1+\frac{\beta}{\sqrt{3}}\right)s_W^2\right]
  \hskip 1cm  (q=u,d,s,c) \, . \label{DLqq} 
  \ee
 Moreover the couplings to right-handed quarks  read
 \bea
\Delta_R^{UU}(Z^\prime)&=&\frac{g }{2\sqrt{3}\, c_W }\sqrt{f(\beta)} \, 
4\frac{\beta}{\sqrt{3}}  s_W^2 \hskip 3.5cm (U=u,c,t)  \,\,\label{DRUU} \\
\Delta_R^{DD}(Z^\prime)&=&-\frac{g }{2\sqrt{3}\, c_W}\sqrt{f(\beta)} \, 2\frac{\beta}{\sqrt{3}} s_W^2 \hskip 3.2cm (D=d,s,b)\,\, .\label{DRDD} 
\eea
Calculating the tree-level $Z^\prime$ exchange diagrams 
we find for both $D\to K^+K^-$ and $D\to \pi^+\pi^-$ decays the  effective Hamiltonian at $M_{Z^\prime}$:
\be
H_{eff}=C_3(M_{Z^\prime}) Q_3+C_7(M_{Z^\prime}) Q_7+\overline C_3(M_{Z^\prime}) \overline Q_3 \, ,
\ee
with the  penguin operators $Q_3$ and $Q_7$ also present  in the SM and defined in Appendix~\ref{Operators}. The new operator
\be
\overline Q_3= (\bar s d)_{V-A}\left[(\bar bb)_{V-A} + (\bar tt)_{V-A}\right]
\ee
arises from the  different couplings of $Z^\prime$ to the third generation of quarks; however,  its contribution, only through RG effects, is negligible because
$b$ and $t$ quarks do not appear in the low energy effective theory relevant
for $D$ decays.  The coefficients $C_{3}$, $C_{7}$  and $\overline C_3$
are expressed in terms of the couplings  in (\ref{DLqq}), (\ref{DRUU}) and (\ref{DRDD}):
\bea
C_3(M_{Z^\prime})&=& \frac{g}{2\sqrt{3}c_W}\sqrt{f(\beta)} \left[-1+(1+\frac{\beta}{\sqrt{3}})s_W^2\right] \frac{\Delta_L^{uc}(Z^\prime)}{4 M_{Z^\prime}^2} \label{C3Zp}\\
C_7(M_{Z^\prime})&=& \frac{g}{2\sqrt{3}c_W}\sqrt{f(\beta)}\frac{4}{\sqrt{3}}\beta s_W^2\frac{\Delta_L^{uc}(Z^\prime)}{4 M_{Z^\prime}^2} \label{C7Zp} \\
\overline C_3(M_{Z^\prime})&=& \frac{g}{2\sqrt{3}c_W}\sqrt{f(\beta)} \left[2 c_W^2\right] \frac{\Delta_L^{uc}(Z^\prime)}{4 M_{Z^\prime}^2}\,\,.
\eea
 The RG evolution to the scale $\mu_c \simeq m_c$ generates other operators. In particular,  the full set of operators with the structure of QCD ($Q_3-Q_6$) and of the EW penguins ($Q_7,\,Q_8$) is generated, as  described in  Appendix \ref{Operators}.
Using short-hand notation for  the various hadronic matrix elements
\be
\langle Q_i\rangle_{KK} \equiv\langle K^+ K^-|Q_i|D^0 \rangle,
\quad
\langle Q_i\rangle_{\pi\pi} \equiv\langle \pi^+ \pi^-|Q_i|D^0 \rangle
 , \ee
the 331 contributions to the decay amplitudes read:
\bea
{\cal A}(D^0 \to K^+ K^-)^{331}&=&\sum_{i=3}^{8}C_i(\mu_c)\langle Q_i\rangle_{KK}
\label{A331K}\\
{\cal A}(D^0 \to \pi^+ \pi^-)^{331}&=&\sum_{i=3}^{8}C_i(\mu_c)\langle Q_i\rangle_{\pi\pi} \, ,
\label{A331pi}
\eea
\noindent
where $\mu_c \simeq {\cal O}(m_c)$. The values of the coefficients $C_i(\mu_c)$ are collected in Appendix~\ref{Operators}. 
We  use such expressions in the numerical analysis in Section \ref{numerics}.

\subsection{$c \to u \ell ^+ \ell^-$ transition:   Wilson coefficients in the effective Hamiltonian and $D^0 \to \mu^+ \mu^-$ }
In 331 models the  tree level $Z^\prime $ mediated FCNC involve only left-handed quarks, while flavour conserving currents involve fermions  of both helicities. Hence, as in the SM
  the effective Hamiltonian for $c \to u \ell ^+ \ell^-$ transition  consists  of two current-current operators:
\be
H_{eff}^{c \to u \ell ^+ \ell^-}={\tilde C}_9(\mu) \, {\tilde Q}_9 +{\tilde C}_{10}(\mu) \, {\tilde Q}_{10}
\ee
with
\be
\label{Hell1}
      {\tilde Q}_9 = \overline{u}_L \gamma^\mu c_L \overline{\ell} \gamma_\mu \ell , \qquad \quad \quad
{\tilde Q}_{10}=\overline{u}_L \gamma^\mu c_L  \overline{\ell} \gamma_\mu \gamma_5\ell\,.
\ee
Neglecting SM contributions, we find for the coefficients in 331:
\be
{\tilde C}_9=\frac{1}{2M_{Z^\prime}^2}\, \Delta_L^{uc}(Z^\prime) \Delta_V^{\ell {\bar \ell}}(Z^\prime),\qquad {\tilde C}_{10}=\frac{1}{2M_{Z^\prime}^2}\, \Delta_L^{uc}(Z^\prime) \Delta_A^{\ell {\bar \ell}}(Z^\prime) , \label{eq:hcull}
\ee
with 
 \bea
\Delta_L^{\ell \bar\ell}(Z') & =& \frac{g~\left[1-(1+\sqrt{3}\beta)s_W^2\right]}{2 \sqrt{3}c_W\sqrt{1-(1+\beta^2)s_W^2}}\,,\label{DLll} \\
\Delta_R^{\ell \bar\ell}(Z')&=&
 \frac{-g~\beta~ s_W^2}{c_W\sqrt{1-(1+\beta^2)s_W^2}} \,,
\label{DRll} 
\eea
and
\be
\Delta_V^{\ell {\bar \ell}}(Z^\prime)=\Delta_R^{\ell {\bar \ell}}(Z^\prime)+\Delta_L^{\ell {\bar \ell}}(Z^\prime),\qquad
\Delta_A^{\ell {\bar \ell}}(Z^\prime)=\Delta_R^{\ell {\bar \ell}}(Z^\prime)-\Delta_L^{\ell {\bar \ell}}(Z^\prime) \,\,.
\ee
Using \eqref{eq:hcull} the $D^0 \to \ell^+ \ell^-$ branching fraction reads:
\be
{\cal B}(D^0 \to \ell^+ \ell^-)= \frac{m_{D^0}}{8 \pi \Gamma_D} \sqrt{1-\frac{4 m_\ell^2}{m_{D^0}^2}}\left|F_D m_\ell {\tilde C}_{10} \right|^2 \label{D0mumu1}.
\ee
In the SM we find, in agreement with  \cite{Fajfer:2015mia}:
  \be
  ( {\tilde C_{10}})_{\rm SM}=\frac{G_F \alpha}{ \sqrt{2}\pi\sin^2\theta_W}V_{cb}^*V_{ub}Y(x_b),\qquad x_b=\frac{m_b^2}{M_W^2}, \label{Ctilde10SM}
  \ee
  where $Y(x_b)$ is the one-loop function that in $B_s\to\mu^+\mu^-$
  appears like $Y(x_t)$ \cite{Buras:2020xsm}. Since $m_b \ll M_W$, we have  to an excellent
  approximation:
  \be
  Y(x_b)=\frac{x_b}{2}\,. \label{Yxb}
  \ee
 The SM prediction is then
\be
{\cal B}(D^0 \to \mu^+ \mu^-)|_{SM}=7.58 \times 10^{-21} \label{eq:D0mumusm}
\ee
in correspondence of the central values for the input parameters.
The present experimental upper bound  is ${\cal B}(D^0 \to \mu^+ \mu^-)<6.2 \times 10^{-9}$ at 90$\%$ C.L.\cite{Zyla:2020zbs}. \footnote{ LD contributions to this decay in the SM include the process  $D^0 \to \gamma \gamma \to \mu^+ \mu^-$, which is however affected by uncertainties, e.g. the  $D^0 \to \gamma \gamma$ rate
\cite{Burdman:2001tf}.}

\section{Numerical results in 331}\label{numerics}
A  strategy for the  numerical analysis of flavour observables in 331 model has been outlined   in \cite{Buras:2015kwd}. The model  parameters are bound imposing that $\Delta M_{B_d},\, S_{J/\psi K_S}$ and $\Delta M_{B_s},\, S_{J/\psi \phi}$ lie in their experimental ranges within  $2\sigma$. In the kaon sector we require   $\epsK  \in [1.6,\,2.5] \times 10^{-3}$ and $\Delta M_K$ varying between $[0.75,\,1.25] \times (\Delta M_K)_{SM}$:  using $V_{ub}$ in Table \ref{tab:input} this corresponds to $(\Delta M_K)_{SM}=0.0047$ ps$^{-1}$.
 We refer to \cite{Buras:2012dp} for the formulae for the various observables in the SM and in 331 models.  We collect  in Table \ref{tab:input} the theory and experimental input parameters  used in the present study. In the case of the CKM matrix,  Table  \ref{tab:input} displays the four entries that we set as independent parameters;  all the other CKM parameters are derived from these ones.
 In particular, $|V_{cb}|$ is fixed at the central value in the Table, while $|V_{ub}|$ and $\gamma$ are chosen within the range quoted in \cite{Zyla:2020zbs}. Among the other CKM parameters, we obtain for $|V_{cd}|$, $|V_{ud}|$ and $|V_{cs}|$  relevant for our study:
$ |V_{cd}|=0.2251$, $|V_{ud}|=0.9743$ and $|V_{cs}|=0.9735$. 
 
\begin{table}[!tb]
\center{\begin{tabular}{|l|l|}
\hline
$G_F = 1.16637(1)\times 10^{-5}\gev^{-2}$\hfill\cite{Zyla:2020zbs} 	&  $m_{B_d}= 5279.63(20)\mev$\hfill\cite{Zyla:2020zbs}\\
$M_W = 80.385(15) \gev$\hfill\cite{Zyla:2020zbs}  								
&	$m_{B_s} =
5366.88(14)\mev$\hfill\cite{Zyla:2020zbs}\\
$\sin^2\theta_W = 0.23121(4)$\hfill\cite{Zyla:2020zbs} 				& 	
$F_{B_d} =
190.0(1.3)\mev$\hfill\cite{Aoki:2019cca}\\
$\alpha(M_Z) = 1/127.9$\hfill\cite{Zyla:2020zbs}									& 	$F_{B_s} =
230.3(1.3)\mev$\hfill\cite{Aoki:2019cca}\\
$\alpha_s^{(5)}(M_Z)= 0.1179(10) $\hfill\cite{Zyla:2020zbs}								&  $\hat B_{B_d} =
1.30(6)$\hfill\cite{Aoki:2019cca}\\\cline{1-1}		
$m_c(m_c) = 1.279(8) \gev$ \hfill\cite{Chetyrkin:2017lif,Zyla:2020zbs}& $\hat B_{B_s} =
1.32(5)$\hfill\cite{Aoki:2019cca}\\
$m_b(m_b)=4.163(16)\gev$\hfill\cite{Chetyrkin:2009fv,Zyla:2020zbs} 			
&
$F_{B_d} \sqrt{\hat
B_{B_d}} = 216(10)\mev$\hfill\cite{Aoki:2019cca}\\$m_t(m_t) = 162.5\pm^{2.1}_{1.5}\gev$\hfill\cite{Zyla:2020zbs}&
 $F_{B_s} \sqrt{\hat B_{B_s}} =
262(10)\mev$\hfill\cite{Aoki:2019cca} 
\\$M_t=172.76(30) \gev$\hfill\cite{Zyla:2020zbs} 	
 &  $\eta_B=0.55(1)$\hfill\cite{Buras:1990fn,Urban:1997gw}  \\
\cline{1-1}
$m_{K^+}= 493.677(13)\mev$	\hfill\cite{Zyla:2020zbs}				& $\Delta M_d = 0.5065(19)
\,\text{ps}^{-1}$\hfill\cite{Zyla:2020zbs}
\\
$m_{K^0}= 497.611(13)\mev$	\hfill\cite{Zyla:2020zbs}		&  $\Delta M_s = 17.756(21)
\,\text{ps}^{-1}$\hfill\cite{Zyla:2020zbs}\\	

$F_K = 156.1(11)\mev$\hfill\cite{Aoki:2019cca}												&
$S_{\psi K_S}= 0.695(19)$\hfill\cite{Zyla:2020zbs}\\
$\hat B_K= 0.7625(97)$\hfill\cite{Aoki:2019cca}												&
$S_{\psi\phi}= 0.054(20)$\hfill\cite{Aoki:2019cca}\\
$\tau(K_L)= 5.116(21)\times 10^{-8}\,\text{s}$\hfill\cite{Zyla:2020zbs}								& $\tau(B_s)= 1.515(4)\,\text{ps}$\hfill\cite{Zyla:2020zbs}\\	
$\tau(K_S)= 0.8954(4)\times 10^{-10}\,\text{s}$\hfill\cite{Zyla:2020zbs}									& $\tau(B_d)= 1.519(4) \,\text{ps}$\hfill\cite{Zyla:2020zbs}\\									\cline{2-2}
$\Delta M_K= 0.5293(9)\times 10^{-2} \,\text{ps}^{-1}$\hfill\cite{Zyla:2020zbs}	&
$|V_{us}|=0.2252(5)$\hfill\cite{Zyla:2020zbs}\\
$|\epsilon_K|= 2.228(11)\times 10^{-3}$\hfill\cite{Zyla:2020zbs}					
& $|V_{cb}|=(41.0\pm1.4)\times
10^{-3}$\hfill\cite{Zyla:2020zbs}\\\cline{1-1}
$m_{D^0} =1864.83(5)
\mev$\hfill\cite{Zyla:2020zbs}
   &
$|V_{ub}|=3.72\times10^{-3}$\hfill\cite{Zyla:2020zbs}\\
$F_D=212.0(7)\mev$\hfill\cite{Aoki:2019cca}
& $\gamma=68^\circ$\hfill\cite{Zyla:2020zbs}
\\	$\tau(D^0)= 0.4101(15)  \,\text{ps}$\hfill\cite{Zyla:2020zbs}		&\\
$B_1=0.82$ \hfill \cite{Gupta:1996yt,Lin:2006vc} & \\
\hline
\end{tabular}  }
\caption {Values of the experimental and theoretical
    quantities used as input parameters  or as  experimental constraints for the observables analyzed in the 331 models.}
\label{tab:input}
\end{table}

 The  allowed ranges for the parameters ${\tilde s}_{13},\,\delta_1$, 
 ${\tilde s}_{23},\, \delta_2$ are shown in Fig.~\ref{oases} for $M_{Z^\prime}\in[ 1,\,  5]\,$ TeV.
The regions in the parameter plane $({\tilde s}_{13},\,\delta_1)$ are obtained imposing the constraints on $\Delta M_{B_d}$ and $S_{J/\psi K_S}$  whose measurements are in Table \ref{tab:input},   the regions for $({\tilde s}_{23},\, \delta_2)$ are obtained using the measured   
$\Delta M_{B_s}$ and $S_{J\psi \phi}$  in the same Table. 
\begin{figure}[!tb]
\begin{center}
\includegraphics[width = 0.8\textwidth]{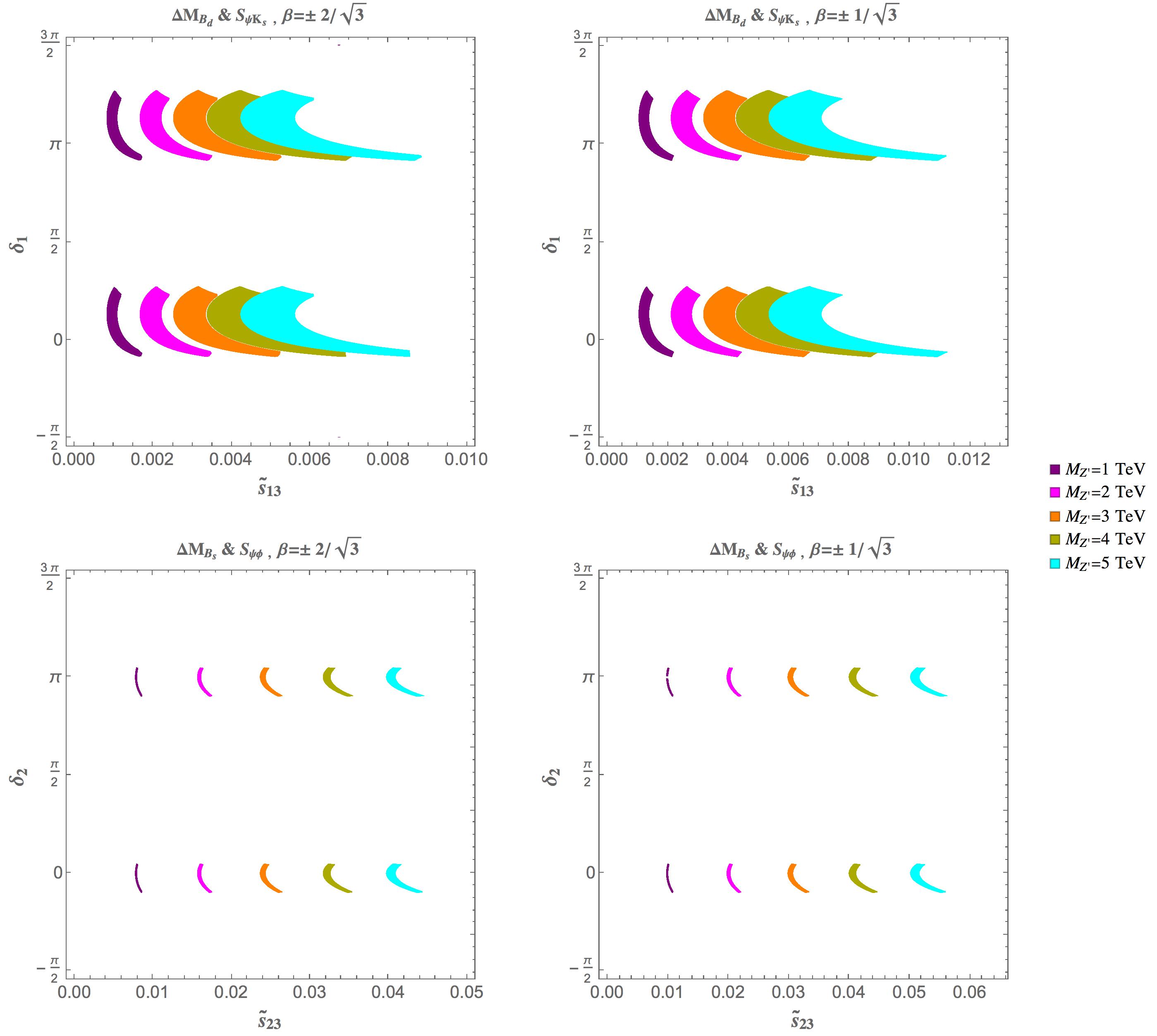}
    \caption{\small Allowed regions in the space of parameters ${\tilde s}_{13},\,\delta_1$, 
 ${\tilde s}_{23},\, \delta_2$, for the models with $\beta=\pm \frac{2}{\sqrt{3}},\, \pm \frac{1}{\sqrt{3}}$,  varying the $Z^\prime$ mass in the range $[1,\,5]$ TeV.}\label{oases}
\end{center}
\end{figure}
All  panels in Fig.~\ref{oases} show  two ranges for the phases $\delta_{1,2}$ which are independent of the value of $M_{Z^\prime}$;   the ranges for the parameters ${\tilde s}_{13(23)}$ depend on $M_{Z^\prime}$.
 
 The observables analyzed in the following are computed varying ${\tilde s}_{13},\,\delta_1$, 
 ${\tilde s}_{23},\, \delta_2$  in their allowed ranges, and selecting only the values for which  the constraints  from $\Delta F=2$ observables in the kaon sector are also satisfied.  

\subsection{$D^0-{\overline D}^0$ mixing}
In Section \ref{charm} the mixing parameters $x_D,\,y_D,\,\left|q/p \right|$ and $\phi$ have been introduced together with the present status of their measurements.
These quantities depend on  $M_{12}^D$ and  $\Gamma_{12}^D$. As already stated, in the SM  $M_{12}^D$ and  $\Gamma_{12}^D$ are real to a very good approximation, hence $\left|q/p \right|^{SM}=1$ and $\phi^{SM}=0$. 

The data in Eq.~(\ref{xDyDexp}) are compatible  with these values for $\,\left|q/p \right|$ and $\phi$, and it is possible to find  the   real values of $(M_{12})^{SM}$ and $(\Gamma_{12})^{SM}$ (the superscript $D$ is omitted)  reproducing  the central values  of  $x_D$ and $y_D$:
\be
\left(({\tilde M}_{12})^{SM},({\tilde \Gamma}_{12})^{SM}  \right)= ( 0.0045 ,\, \,  0.0166) \, {\rm ps}^{-1}.
\label{M12G12SM}
\ee
The 331 model provides  a new contribution to $M_{12}^D$ with  both real and imaginary part.  Therefore, we can write the contribution of SM and 331 in the form:
\bea
{\rm Re}[M_{12}^D]&=&(M_{12})^{SM}+{\rm Re}[(M_{12})^{331}] \, ,
\hskip 1cm  {\rm Im}[M_{12}^D]={\rm Im}[(M_{12})^{331}] \, ,
\label{M12tot}
\\
{\rm  Re}[\Gamma_{12}^D]&=&(\Gamma_{12})^{SM} \, , \hskip 3.8 cm {\rm Im}[\Gamma_{12}^D]=0 \, .
\label{Gamma12tot}
\eea
The  331 contribution shifts the SM terms $(M_{12})^{SM},\,(\Gamma_{12})^{SM}$ from $({\tilde M}_{12})^{SM}$, $({\tilde \Gamma}_{12})^{SM}$.
%
To investigate the impact of the 331 model we vary $(M_{12})^{SM}$ and $(\Gamma_{12})^{SM}$ in a range centered at the values
$\left(({\tilde M}_{12})^{SM},({\tilde \Gamma}_{12})^{SM}  \right)$.  We
select the set of parameters $({M}_{12})^{SM}$, $({\Gamma}_{12})^{SM}$, ${\tilde s}_{13},\,\delta_1,\,{\tilde s}_{23},\,\delta_2$ for which    the computed $x_D$ is in the experimental range within $1\sigma$  and $2\sigma$.
In correspondence to such sets of values we determine $\left|q/p \right| $ and $\phi$, providing  an interesting correlation between $\phi$ and the function $(1-|q/p |)$ which measures the deviation of $|q/p |$ from unity.
The result is depicted in Fig.~\ref{qsup}, with
the magenta (light blue)  regions  corresponding to the set of parameters  for which the sum LD SM + 331  reproduce $x_D$ within $1\sigma$ ($2\sigma$). 
Without any additional contribution, the 331 would not be able to reproduce the experimental $x_D$: indeed, the expressions \eqref{xD331},\eqref{xD331evol} result in a value for  $x_D^{331}$ in the range $(0.2-2.5)\times 10^{-4}$.

It is interesting to consider the correlation between the asymmetry  $S_f$ (for $\eta_f=+1$) and the semileptonic asymmetry  $a_{SL}(D^0)$ in Eqs.~\eqref{eq:Sf},\eqref{eq:ASL}. It is depicted in Fig.~\ref{ASLSF} for $\xi_f=0$.  The result does not change if the phase $\xi_f$ obtained in correspondence to the parameters fixed in the analysis of $\Delta A_{CP}$ described below,  are included in the expression \eqref{eq:CORR2}. A semileptonic asymmetry $a_{SL}$ at the per-cent level is permitted, while $|S_f|$ does not exceed ${\cal O}(10^{-4})$.

\begin{figure}[t]
\begin{center}
\includegraphics[width = 0.8\textwidth]{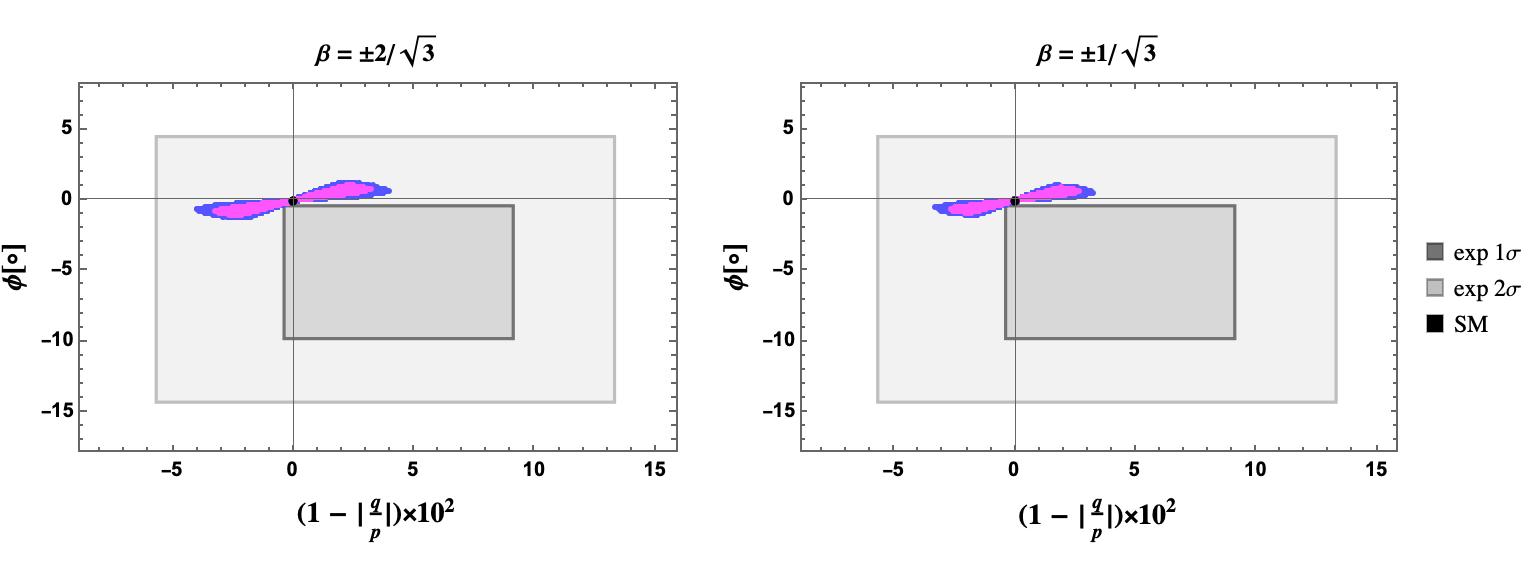}
    \caption{\small Correlation between the phase $\phi$ (in degrees) in $D^0 - {\overline D}^0$ mixing \eqref{qoverp} and the function $(1-|q/p|)$ measuring the deviation of $|q/p|$ from unity, obtained in the 331 models for different values of the parameter $\beta$ and for $M_{Z^\prime}=1$ TeV. The magenta (blue) region is obtained imposing that  the mixing observables $x_D,\,y_D$ lie in the experimental range in \eqref{xDyDexp} within 1$\sigma$ (2$\sigma$). The gray (light gray) regions show  the 1$\sigma$ (2$\sigma$) range for $(\phi,\,|q/p|)$ in Eq.~\eqref{xDyDexp}.} \label{qsup}
\end{center}
\end{figure}
\begin{figure}[!b]
\begin{center}
\includegraphics[width = 0.8\textwidth]{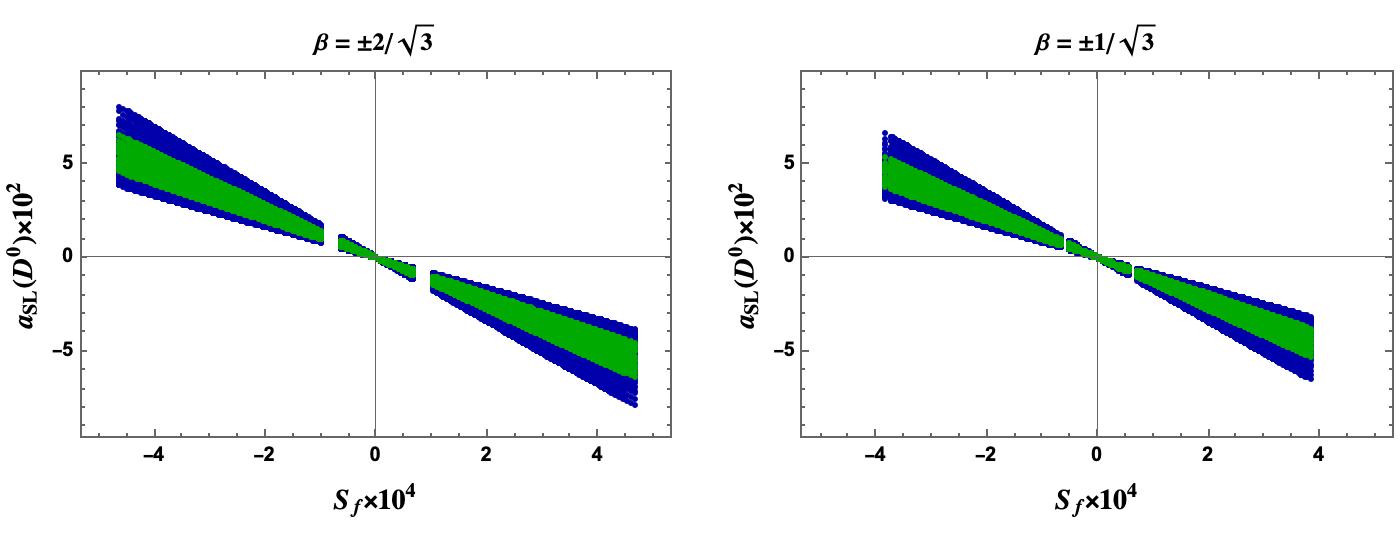}
    \caption{\small    Correlation between the asymmetry  $S_f$ (for $\eta_f=+1$) and the semileptonic asymmetry  $a_{SL}(D^0)$  Eqs.~\eqref{eq:Sf},\eqref{eq:ASL} in 331 models with parameters as in Fig.~\ref{qsup}. The green (blue) region corresponds to  the mixing observables $x_D$ and $y_D$  within 1$\sigma$ (2$\sigma$) in the experimental range  in \eqref{xDyDexp}. } \label{ASLSF}
\end{center}
\end{figure}
\subsection{$\Delta A_{CP}$}

 The 331 model produces a new contribution to the CP asymmetry (\ref{ACP}) for $f=K^+ K^-$ and $f=\pi^+ \pi^-$,
\be
({A}_{CP}^f)^{331}=-2 {\tilde r}_f \, \sin{\tilde \delta}_f \, \sin {\tilde \phi}_f \label{ACPNP} \,\,,
\ee
where
\bea
{\tilde r}_{K^+ K^-}&=&\left|\frac{{\cal A}(D^0 \to K^+ K^-)^{331}}{{\cal A}_{SM,K}^{tree}}\right| \, ,
\quad \hspace*{0.1cm} {\tilde \phi}_{K^+ K^-}={\rm Arg}\left[\frac{{\cal A}(D^0 \to K^+ K^-)^{331}}{{\cal A}_{SM,K}^{tree}}\right] \, , \quad \label{phitildeK}
\\
{\tilde r}_{\pi^+ \pi^-}&=&\left|\frac{{\cal A}(D^0 \to \pi^+\pi^-)^{331}}{{\cal A}_{SM,\pi}^{tree}}\right| \, , 
\quad \quad {\tilde \phi}_{\pi^+ \pi^-}={\rm Arg}\left[\frac{{\cal A}(D^0 \to \pi^+ \pi^-)^{331}}{{\cal A}_{SM,\pi}^{tree}}\right] \, , \label{phitildepi}
\eea
and ${\cal A}(D^0 \to K^+ K^- (\pi^+ \pi^-))^{331}$  in (\ref{A331K}),(\ref{A331pi}).
In our analysis   we assume a maximal  difference $\tilde \delta_f$ for the strong phases. 
The tree-level SM contributions to the $D^0 \to K^+ K^- (\pi^+\pi^-)$ amplitudes read:
\bea
{\cal A}_{SM,K}^{tree}=\lambda_s \, {\cal A}_{SM,K}^{T}= \lambda_s \sum_{i=1}^2 C_i(\mu_c) \langle Q_i \rangle_{KK} \label{ASMtreeK} \\
{\cal A}_{SM,\pi}^{tree}=\lambda_d \, {\cal A}_{SM,\pi}^{T}= \lambda_d \sum_{i=1}^2 C_i(\mu_c) \langle Q_i \rangle_{\pi \pi} \, , \label{ASMtreepi} 
\eea
where  $C_1(m_c)=1.2835$ and $C_2(m_c)=-0.5467$. 

The main uncertainty in the amplitudes  \eqref{ASMtreeK},\eqref{ASMtreepi} and \eqref{A331K},\eqref{A331pi} is related to the value of  the hadronic matrix elements $\langle Q_i \rangle_{KK}$ and  $\langle Q_i \rangle_{\pi \pi}$. 
The simplest approach to evaluate such quantities,  the naive factorization (NF) prescription,
  allows to express the matrix elements of the relevant operators in terms of  $\langle Q_1 \rangle_f$:
\bea
\langle Q_1 \rangle_f&=&N_c \langle Q_2 \rangle_f=N_c \langle Q_3 \rangle_f= \langle Q_4  \rangle_f \nn \\
\chi_f  \langle Q_1 \rangle_f&=&N_c \langle Q_5 \rangle_f= \langle Q_6 \rangle_f=-2N_c \langle Q_7 \rangle_f=-2 \langle Q_8 \rangle_f \,\,, \label{naive}
\eea
with $N_c$ the number of colors.
In (\ref{naive}) the chiral factors are  $\chi_{K^+K^-}=\displaystyle\frac{2m_{K^+}^2}{m_c m_s} \simeq 4.06$ and $\chi_{\pi^+{\pi^-}}=\displaystyle\frac{2m_{\pi^+}^2}{m_c (m_u+m_d)} \simeq 4.45$. The 331 contribution to $\Delta A_{CP}$ can be worked out
using these relations and varying the 331 parameters in  the allowed oases.
 The largest value is obtained for $\beta=-2/\sqrt{3}$:
$
|(\Delta A_{CP}^{331})_{max}| \simeq 5.6 \times 10^{-5} .
$
The prediction for the sign cannot be provided without an information on the  strong
phases.

However, the accuracy of naive factorization is doubtful,  in particular for charm. On the other hand, no information on  the hadronic  matrix elements is available from QCD methods as lattice QCD.  For this reason, we follow a different  approach.  
We consider the ratios 
\be
\qquad  R_i^f=\displaystyle\frac{\langle Q_i \rangle_f}{ \langle Q_1 \rangle_f} , \qquad \qquad i=2, \dots 8  , \quad f=\pi^+ \pi^-,\,K^+K^- \label{eq:opratios}
\ee 
 as additional quantities to be constrained using data.  
In the numerical analysis, together with the  331 parameters ${\tilde s}_{13},\,{\tilde s}_{23},\,\delta_1,\,\delta_2$, we scan the space of  $R_i^f$ around the  values corresponding to (\ref{naive}), in large intervals.
For example,   $R_{2}^f$ is scanned in the range $[0.33,\,0.83]$,  $R_{3,5,7}^f$ in a range of width $10$, $R_{4}^f$ in a range of width $15$,  $R_{6,8}^f$ in a range of width $30$, allowing large values of the hadronic matrix elements.
The scans are constrained by the 
 experimental  $D^0 \to K^+ K^-$ and $D^0 \to \pi^+ \pi^-$ CP asymmetries  \cite{Amhis:2019ckw} \footnote{The two individual asymmetries in \eqref{ACPKK}, \eqref{ACPpipi}  are affected by sizable uncertainties, however their difference  \eqref{deltaACPexp} is  measured with  higher accuracy.}:
\bea
{A}_{CP}^{K^+ K^-}&=&(-1.33 \pm 1.37) \times 10^{-3} \label{ACPKK}
\\
{A}_{CP}^{\pi^+ \pi^-}&=&(0.27 \pm 1.37) \times 10^{-3} \label{ACPpipi}\,\,,
\eea
which are imposed to be reproduced within $2\sigma$ around the central value.
The  resulting ratios   $\tilde r_f$  and phases $\tilde \phi_f$  in Eq.~(\ref{ACPNP}), (\ref{phitildeK}), (\ref{phitildepi}), are  in Figs.~\ref{ACPK} and \ref{ACPpi}.  
The phases are different from zero, the ratios $\tilde r_f$  span a range up to fractions of  $10^{-3}$.  %
\begin{figure}[!t]
\begin{center}
\includegraphics[width = 0.8\textwidth]{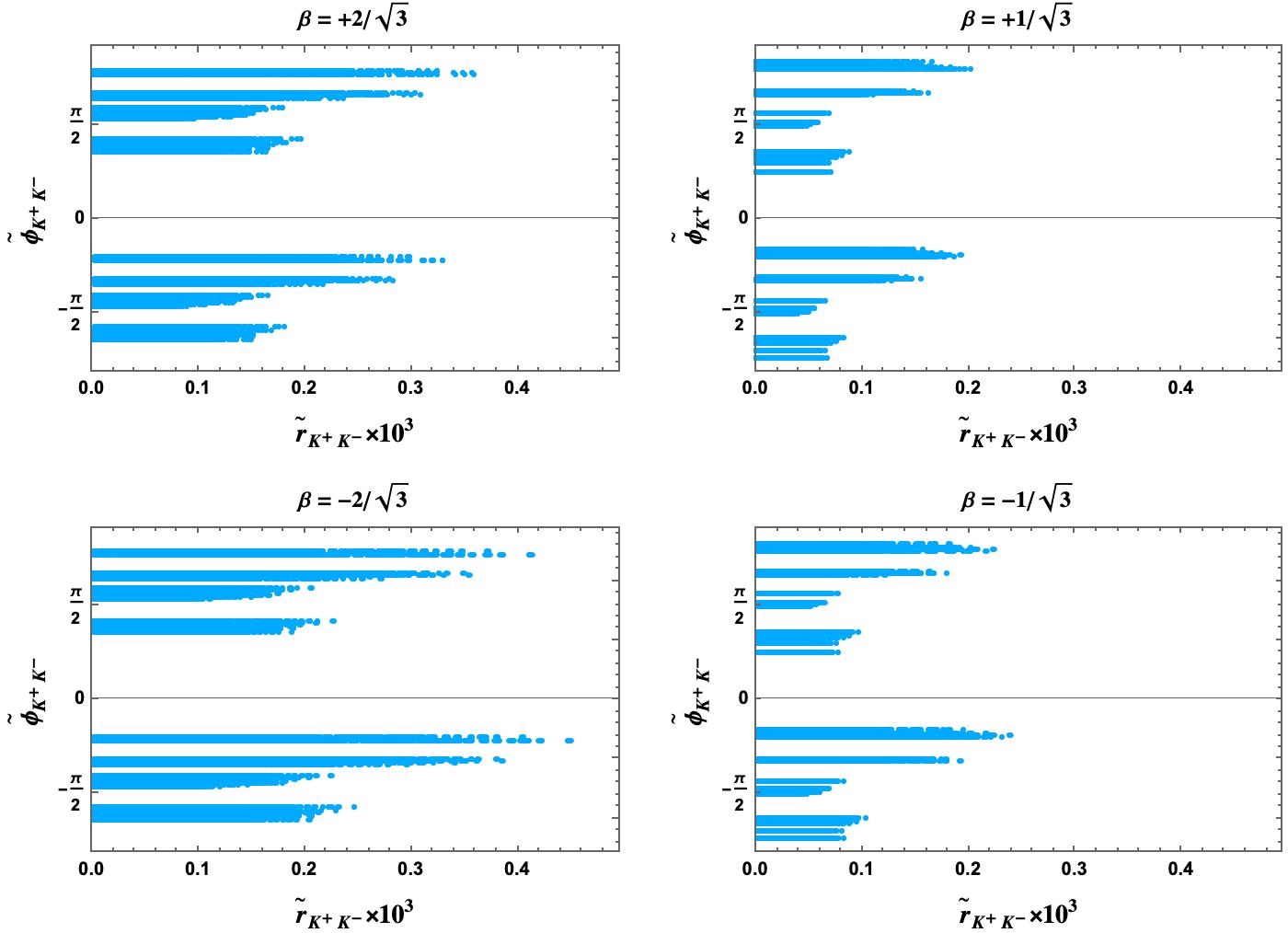}
    \caption{\small Correlation between the ratio ${\tilde r}_{K^+ K^-}$  and  phase ${\tilde \phi}_{K^+ K^-}$
in \eqref{phitildeK}     in  331 models with the same parameters of Fig.~\ref{qsup},  varying the ratios of the operator matrix elements in \eqref{eq:opratios} as described in the text.}  \label{ACPK}
\end{center}
\end{figure}
\begin{figure}[!bt]
\begin{center}
\includegraphics[width = 0.8\textwidth]{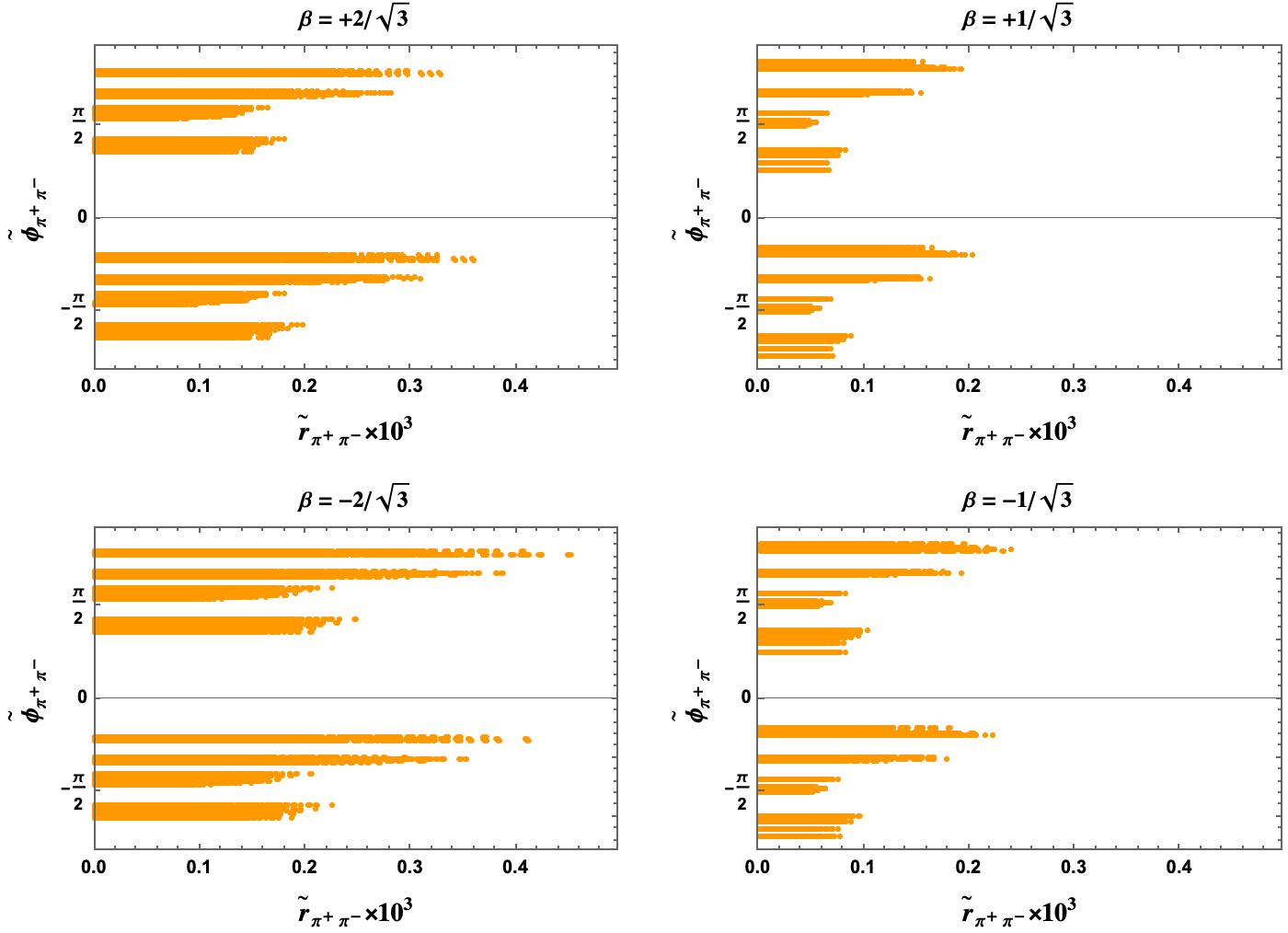}
    \caption{\small Correlation between the ratio ${\tilde r}_{\pi^+ \pi^-}$  
and phase ${\tilde \phi}_{\pi^+ \pi^-}$ in (\ref{phitildepi})  in  331 models, as described in the text.} \label{ACPpi}
\end{center}
\end{figure}
\begin{figure}[!b]
\begin{center}
\includegraphics[width = 0.6\textwidth]{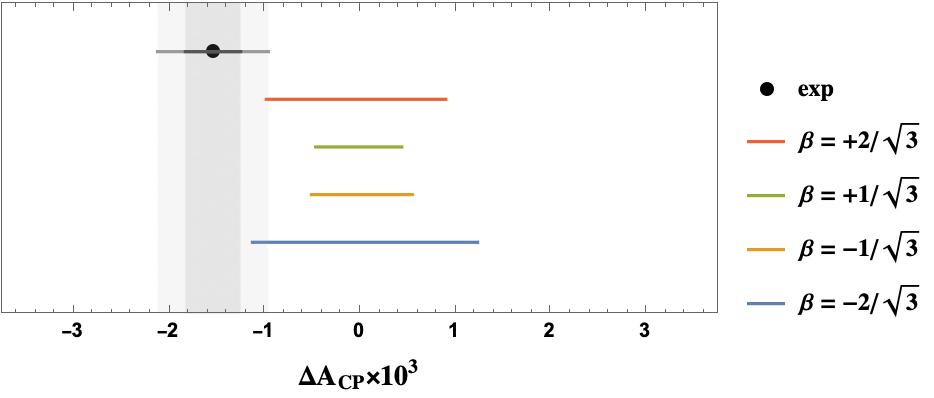}
    \caption{\small Range for $\Delta A_{CP}^{331}$  in 331 models.  $\Delta A_{CP}^{331}$ is obtained for each $\beta$  using the ratios
    ${\tilde r}_{K^+ K^-}$,  ${\tilde r}_{\pi^+ \pi^-}$  
and  phases ${\tilde \phi}_{K^+ K^-}$,  ${\tilde \phi}_{\pi^+ \pi^-}$   in Figs.~\ref{ACPK}, \ref{ACPpi}.  
The  black dot and  bar  correspond to the measurement   in Eq.~\eqref{deltaACPexp}.
    } \label{deltaACP}
\end{center}
\end{figure}
%
Large values of $\tilde r_f$ correspond to large values of the hadronic matrix elements, a  
statement that can be made more precise if the difference $\Delta A_{CP}^{331}$ is considered, as in  
Fig.~\ref{deltaACP}.
When the ratios $R_i^f$ are varied in the ranges quoted above,  for the variants with $\beta=\pm 2/\sqrt{3}$ it is possible to find a set of parameters (the 331 parameters plus ratios of hadronic matrix elements) giving $\Delta A_{CP}$ in agreement with data.
For larger values of the $R_i^f$   also for $\beta=\pm 1/\sqrt{3}$ $\Delta A_{CP}$ can be obtained.
An insight into the required size of the ratios of the hadronic matrix elements can be gained considering  $\beta=-2/\sqrt{3}$ for which, as shown in Fig.~\ref{deltaACP}, the 331 result for $\Delta A_{CP}$ has an overlap with the experimental range. 
The largest value of $|\Delta A_{CP}|$  is obtained in correspondence of  the ratios $R_i^{K^+K^-}$:
 \bea
 R_2^{K^+K^-}&=&0.83 ,  \quad \quad
 R_3^{K^+K^-}=-9.7, \quad
 \nn \\
 R_4^{K^+K^-}&=&+15.5  , \quad
 R_5^{K^+K^-}=-8.7, \quad
 R_6^{K^+K^-}=34 , \nn \\
 R_7^{K^+K^-}&=&-10.7, \quad \quad
 R_8^{K^+K^-}=-32 \,\, , \label{R7K} 
 \eea
 and   of the ratios  $R_i^{\pi^+\pi^-}$:
 \bea
 R_2^{\pi^+\pi^-}&=&0.83 ,  \quad \quad
 R_3^{\pi^+\pi^-}=-9.7, \quad
 \nn \\
 R_4^{\pi^+\pi^-}&=&15.5  , \quad
 R_5^{\pi^+\pi^-}=-8.5, \quad
 R_6^{\pi^+\pi^-}=34.5 , \nn \\
 R_7^{\pi^+\pi^-}&=&-10.7 , \quad
 R_8^{\pi^+\pi^-}=-32.2 \,\, . \label{R7pi} 
 \eea
The conclusion is that large  individual CP asymmetries and $\Delta A_{CP}$ can be obtained provided that  two conditions are  verified:  the phases are in the ranges obtained  in 331 models and the ratios of the hadronic matrix elements are large, close to the values in  \eqref{R7K} and \eqref{R7pi}. This last condition can be verified by  future explicit calculations, namely using lattice QCD.


\subsection{$D^0 \to \mu^+ \mu^-$,  correlations with $B$ and $K$ observables}
We shall next investigate correlations of charm observables with those in the
$B$ and $K$ systems,  a striking possibility within the 331 models. 

In Fig.~\ref{plSBSf} we show the  correlation  between the $S_f$ asymmetry in (\ref{eq:Sf})  and the mixing-induced CP asymmetries  
    $S_{J/\psi K_S}$ and $S_{J/\psi \phi}$ 
in the $B_{d}$  and $B_{s}$ systems.  The message from these plots is clear.
CP violation in the charm system is by orders of magnitude smaller than 
in the $B_{d}$  and $B_{s}$ systems. Even in the $B_s$ system $S_f$ is roughly by
two orders of magnitude smaller than $S_{J/\psi \phi}$.

\begin{figure}[!tb]
\begin{center}
\includegraphics[width = 0.9\textwidth]{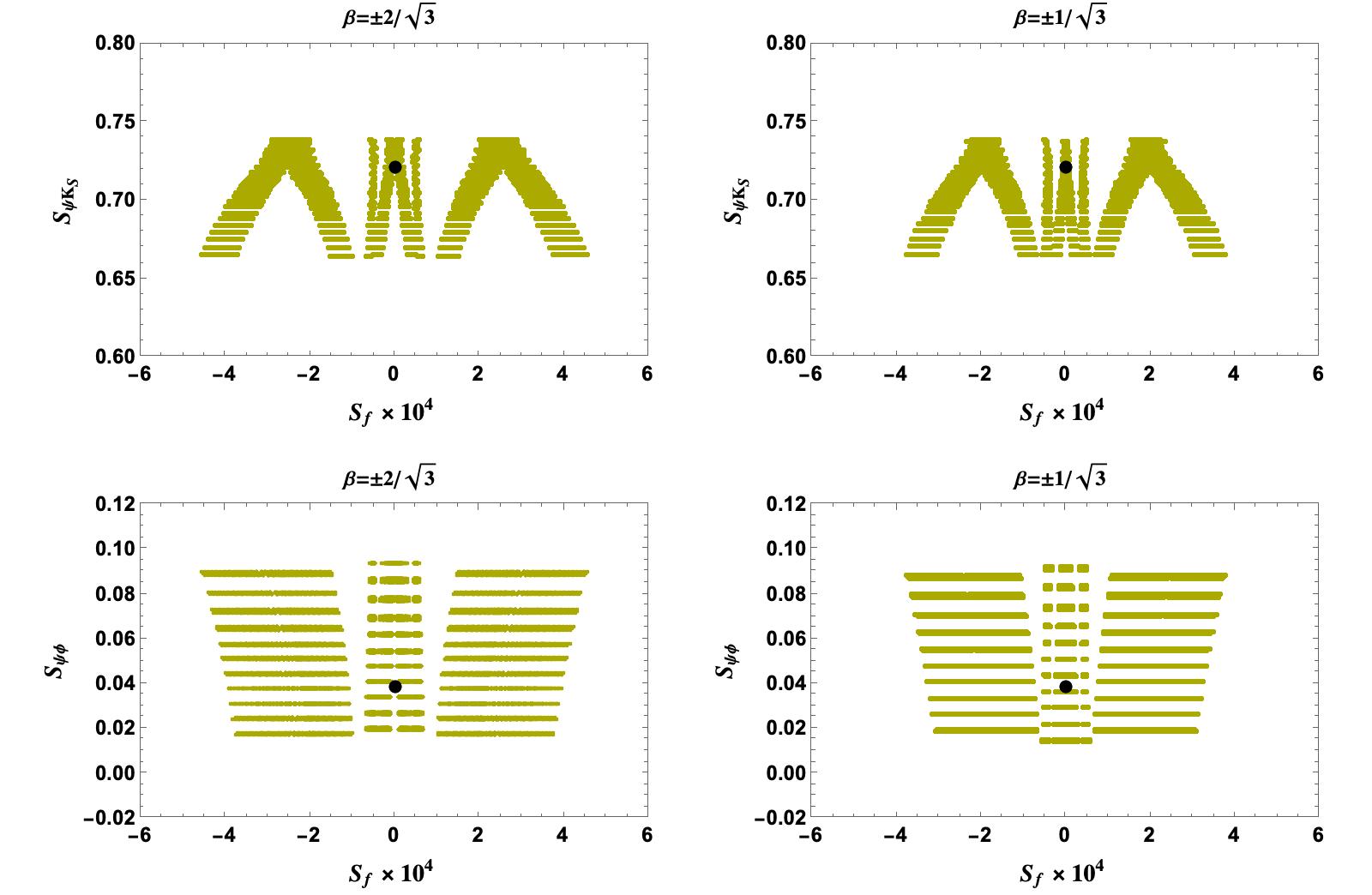}
    \caption{\small Correlation in the  331 models between the $S_f$ asymmetry in (\ref{eq:Sf})  and the CP asymmetries  
    $S_{J/\psi K_S}$ and $S_{J/\psi \phi}$ 
    in the $B_{d}$  and $B_{s}$ systems.   The 331 parameters are as  in Fig.\ref{qsup}. The black dot is the SM result. }\label{plSBSf}
\end{center}
\end{figure}

In Fig.~\ref{plD0xD} we show the correlation between $x_D^{331}$ and ${\cal B}(D^0 \to \mu^+ \mu^-)$ in 331 for the four variants with $\beta$ considered in this paper and $M_{Z^\prime}=1$ TeV. The SM contribution is set to zero in both observables.
We note that ${\cal B}(D^0 \to \mu^+ \mu^-)$ can reach values of order $10^{-14}$,
that is by roughly six orders of magnitude larger than the result \eqref{eq:D0mumusm} obtained within the SM but
still much smaller than the experimental bound. 

 It is worth considering the impact on these results of the measurement 
\be
{\cal {\overline B}}(B_s \to \mu^+ \mu^-)=(2.85 \pm^{0.34}_{0.31}) \times 10^{-9} \,,
\label{BSmumuexp}
\ee
where 
$\overline {\cal B}(B_s \to \mu^+ \mu^-)$ is the average time integrated branching ratio  \cite{DeBruyn:2012wk}:
\be
{\cal {\overline B}}(B_s \to \mu^+ \mu^-)={\cal B}(B_s \to \mu^+ \mu^-)\frac{1+{\cal A}_{\Delta \Gamma}\, y_s}{1-y_s^2} \,\, , \label{BRbar}
\ee
with
$\displaystyle{
{\cal A}_{\Delta \Gamma} = 2\frac{ {\rm Re}[\lambda_{\mu^+ \mu^-}]}{1+|\lambda_{\mu^+ \mu^-}|^2} }$
 and
$
y_s =\displaystyle \frac{\tau_{B_s} \Delta \Gamma_s}{2}=0.0064 \pm 0.004$  \cite{Zyla:2020zbs}.
$\lambda_{\mu^+ \mu^-}$ is analogous  to $\lambda_f$ in Eq.~(\ref{eq:lf}) for $B_s \to \mu^+ \mu^-$, $\tau_{B_s}$ is the $B_s$ mean lifetime and $\Delta \Gamma_s$ the width difference of the heavy and light $B_s$ mass eigenstates.
In Fig.~\ref{plD0xD},  
 imposing  that $\overline {\cal B}(B_s \to \mu^+ \mu^-)$ lies in the experimental range in Eq.~(\ref{BSmumuexp}) within $2\sigma$ we find the blue regions which show that this constraint allows
 only for values of ${\cal B}(D^0 \to \mu^+ \mu^-)$ below $10^{-14}$.
The reason for this further suppression is evident from  Fig.~\ref{plD0Bs} where we show the correlation between $\overline{\cal B}(B_s \to \mu^+ \mu^-)$ and ${\cal B}(D^0 \to \mu^+ \mu^-)$ in 331 models with parameters as in Fig.\ref{plD0xD}.
Indeed,  to suppress the SM branching ratio for $B_s \to \mu^+ \mu^-$ to achieve a better agreement with the experimental measurement
(\ref{BSmumuexp}),
a smaller ${\cal B}(D^0 \to \mu^+ \mu^-)$ is implied. This would not be the case if the data for 
$\overline{\cal B}(B_s \to \mu^+ \mu^-)$ was above the SM expectation.

In Figs.~\ref{plkpiuD0mumu} and \ref{plk0D0mumu} we show the correlations of 
${\cal B}(D^0 \to \mu^+ \mu^-)$ with the branching ratios for 
$K^+ \to \pi^+  \nu \bar \nu$ and $K_L \to \pi^0  \nu \bar \nu$, respectively.
We observe that the largest values of ${\cal B}(D^0 \to \mu^+ \mu^-)$ are found
for the smallest 331 contributions to both kaon decays. The largest contributions to these decays in 331 models amount to roughly $\pm 10\%$ of the SM value
and these modest effects could be tested one day, in particular if the future
data turned out to depart significantly from the SM predictions within the SM
central value represented by black dots in these figures.

The pattern turns out to be different in Figs.~\ref{plkLmumuD0mumu} and
\ref{plkSmumuD0mumu}, where we show the correlations of ${\cal B}(D^0 \to \mu^+ \mu^-)$
with branching ratios  for $K_L\to\mu^+\mu^-$ and  $K_S\to\mu^+\mu^-$, respectively.  Without the ${\cal {\overline B}}(B_s \to \mu^+ \mu^-)$ constraint the
largest 331 contributions to  $K_L\to\mu^+\mu^-$ and  $K_S\to\mu^+\mu^-$ are
accompanied by largest contributions to ${\cal B}(D^0 \to \mu^+ \mu^-)$.
This is in particular seen in the case of $K_L\to\mu^+\mu^-$ in 
 Fig.~\ref{plkLmumuD0mumu}.
However, when the data on ${\cal {\overline B}}(B_s \to \mu^+ \mu^-)$ are
imposed, 331 effects in $D^0 \to \mu^+ \mu^-$
are dwarfed as already seen in Fig.~\ref{plD0Bs}.

\begin{figure}[!tb]
\begin{center}
\includegraphics[width = 0.9\textwidth]{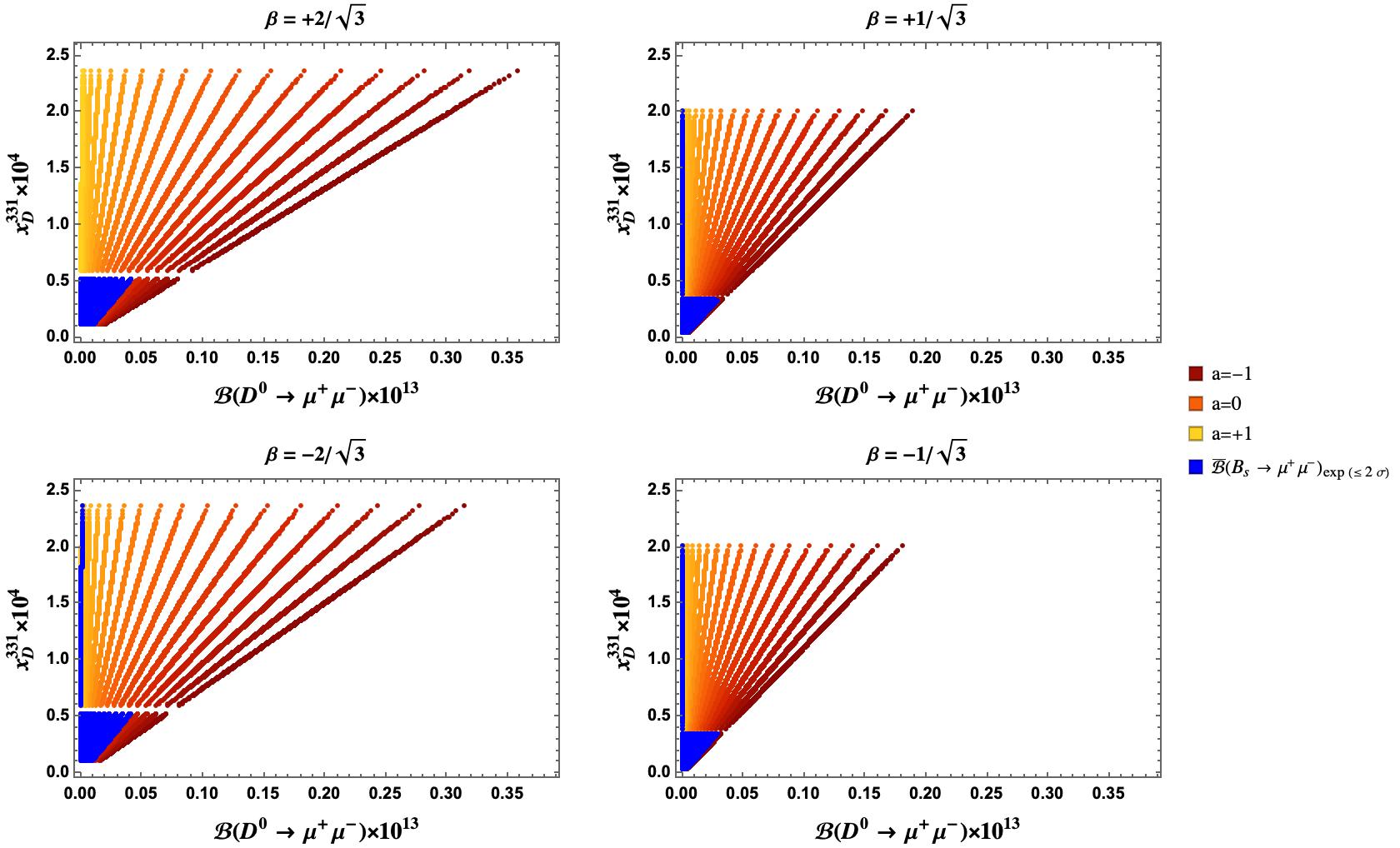}
    \caption{\small Correlation between $x_D^{331}$ and ${\cal B}(D^0 \to \mu^+ \mu^-)$ in the four 331 variants considered in this paper, with $M_{Z^\prime}=1$ TeV. The SM contribution is set to zero in both observables.
     The sliding colours correspond to the variation of  the  $Z-Z^\prime$ mixing parameter $a$ in  Eq.~\eqref{sxi} in the range $ [-1,1]$.    Three of such values are indicated  in the legends. The blue points are obtained imposing that $\overline {\cal B}(B_s \to \mu^+ \mu^-)$ lies in the experimental range within $2\sigma$.}\label{plD0xD}
\end{center}
\end{figure}
\begin{figure}[!tb]
\begin{center}
\includegraphics[width = 0.9\textwidth]{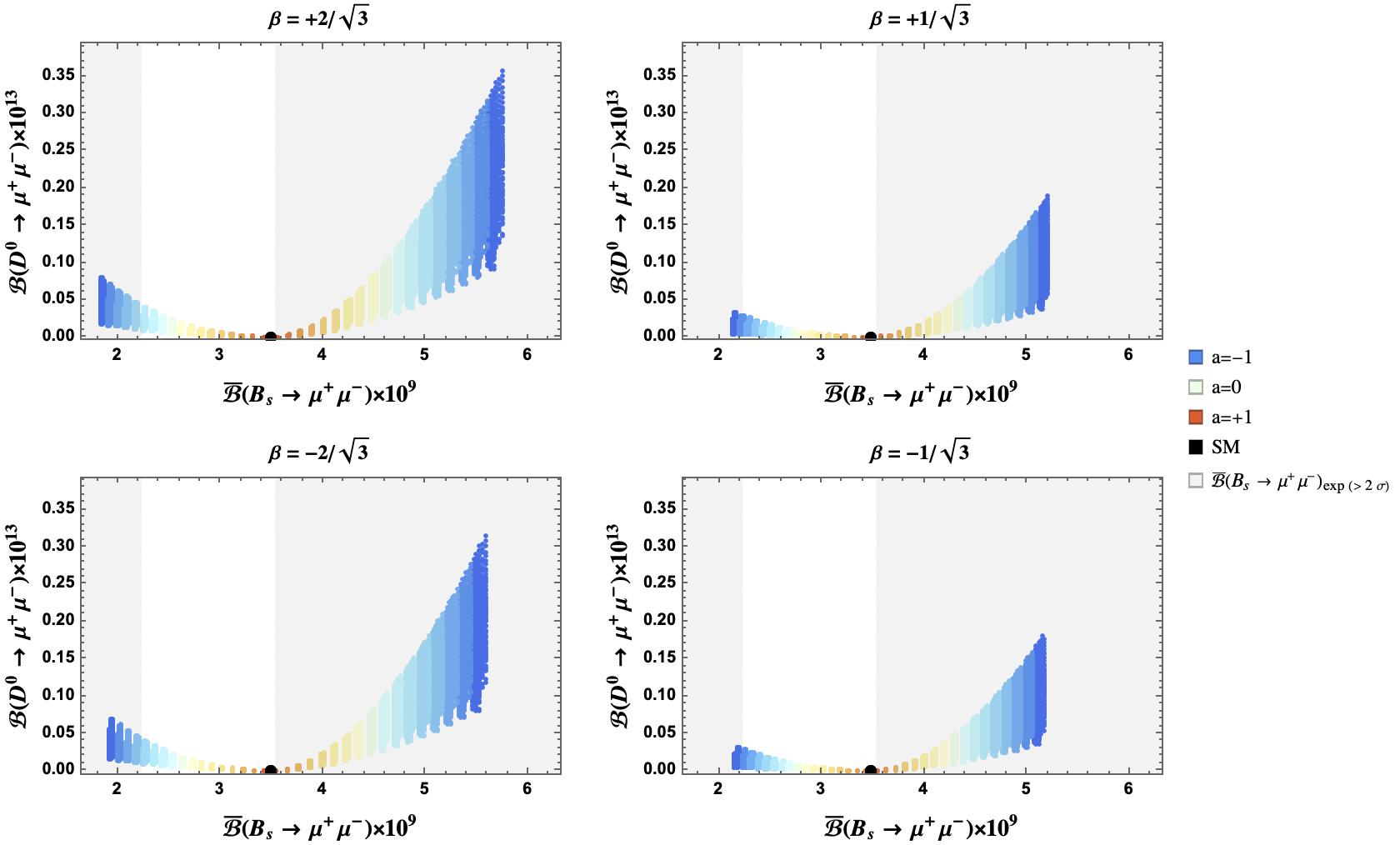}
    \caption{\small Correlation between ${\cal B}(D^0 \to \mu^+ \mu^-)$ and $\overline {\cal B}(B_s \to \mu^+ \mu^-)$  in 331 models with parameters as in Fig.~\ref{plD0xD}.
     The black dot represents the SM result. The shaded gray areas correspond to regions excluded by the average of the experimental measurements of ${\cal {\overline B}}(B_s \to \mu^+ \mu^-)$ within 2$\sigma$ in Eq.~(\ref{BSmumuexp}).}\label{plD0Bs}
\end{center}
\end{figure}

\section{Conclusions}
We have performed a detailed analysis of FCNC processes
in the charm sector in the context four variants of 331 models. The motivation for this analysis arose from the observation \cite{Colangelo:2021myn} that in the case of $Z^\prime$ contributions there are no new free parameters beyond those present already in the
$B_{d,s}$ and $K$ meson systems. As a result definite ranges for NP effects
in various charm observables could be obtained.

In the SM FCNC in the charm system are very strongly GIM suppressed so that
a contribution of $Z^\prime$ that appears already at tree-level could in principle imply NP effects in the charm sector by orders of magnitude larger than
it is possible within the SM.  In fact if we performed our analysis first
for the charm system, ignoring correlations with $B_{d,s}$ and $K$ meson systems,
very large effects would be possible. However, as our analysis demonstrates
most of these effects would be subsequently dwarfed by imposing the contraints
from $B_{d,s}$ and $K$ meson systems that were analyzed in the past \cite{Buras:2012dp,Buras:2013dea,Buras:2014yna,Buras:2015kwd,Buras:2016dxz}.

In particular we find very small effects in $D^0-\bar D^0$ mixing although
the effects are still larger than short-distance contributions within the SM.

Our results are presented in the plots in Section~\ref{numerics} and in the accompanying 
comments. Here we list the most remarkable results.

\begin{itemize}
\item
  As seen in Fig.~\ref{ASLSF} the semileptonic asymmetry  $a_{\rm SL}(D^0)$ can
  be as large as $\pm 5\times 10^{-2}$ and therefore much larger than in the SM.
\item
  As seen in Fig.~\ref{plD0Bs}, the correlation between $\overline{\cal B}(B_s \to \mu^+ \mu^-)$ and ${\cal B}(D^0 \to \mu^+ \mu^-)$ implies the suppression of the latter
  branching ratio if a better agreement with the experimental data
for $\overline{\cal B}(B_s \to \mu^+ \mu^-)$ is to be achieved. Yet, despite this suppression the branching ratio ${\cal B}(D^0 \to \mu^+ \mu^-)$ can reach values of a few $10^{-15}$ which is by six orders larger than its SM value.
\item
  The correlations between  ${\cal B}(D^0 \to \mu^+ \mu^-)$ and the branching ratios for  $\kpn$, $\klpn$ and  $K_{L,S}\to\mu^+\mu^-$ show patterns that depend on the specific 331 model considered. However, if future data on theoretically
  clean decays $\kpn$ and $\klpn$ and also short distance part in $K_{S}\to\mu^+\mu^-$ will show deviations from SM expectations by more than $15\%$ the 331  models will not be able to explain them.
  \item
The $D^0 \to \pi^+ \pi^-, K^+ K^-$  CP asymmetries and $\Delta A_{CP}$ can be sizable for  phases  in the ranges provided in 331 model, however  the ratios of the matrix elements  in Eq.~\eqref{eq:opratios}
  should be large, with a size as provided in \eqref{R7K},\eqref{R7pi}. This possibility  requires a nonperturbative calculation of the matrix elements.
\end{itemize}

We are looking forward to the improved experimental data on all observables
discussed by us to see whether 331 models will survive these tests.

\section*{Acknowledgements}
We thank S. Fajfer and S. Schacht for discussions.
This study has been  carried out within the INFN project (Iniziativa Specifica) QFT-HEP.
A.J.B acknowledges financial support from the Excellence Cluster ORIGINS,
funded by the Deutsche Forschungsgemeinschaft (DFG, German Research Foundation)
under Germany´s Excellence Strategy – EXC-2094 – 390783311.

\begin{figure}[!b]
\begin{center}
\includegraphics[width = 0.9\textwidth]{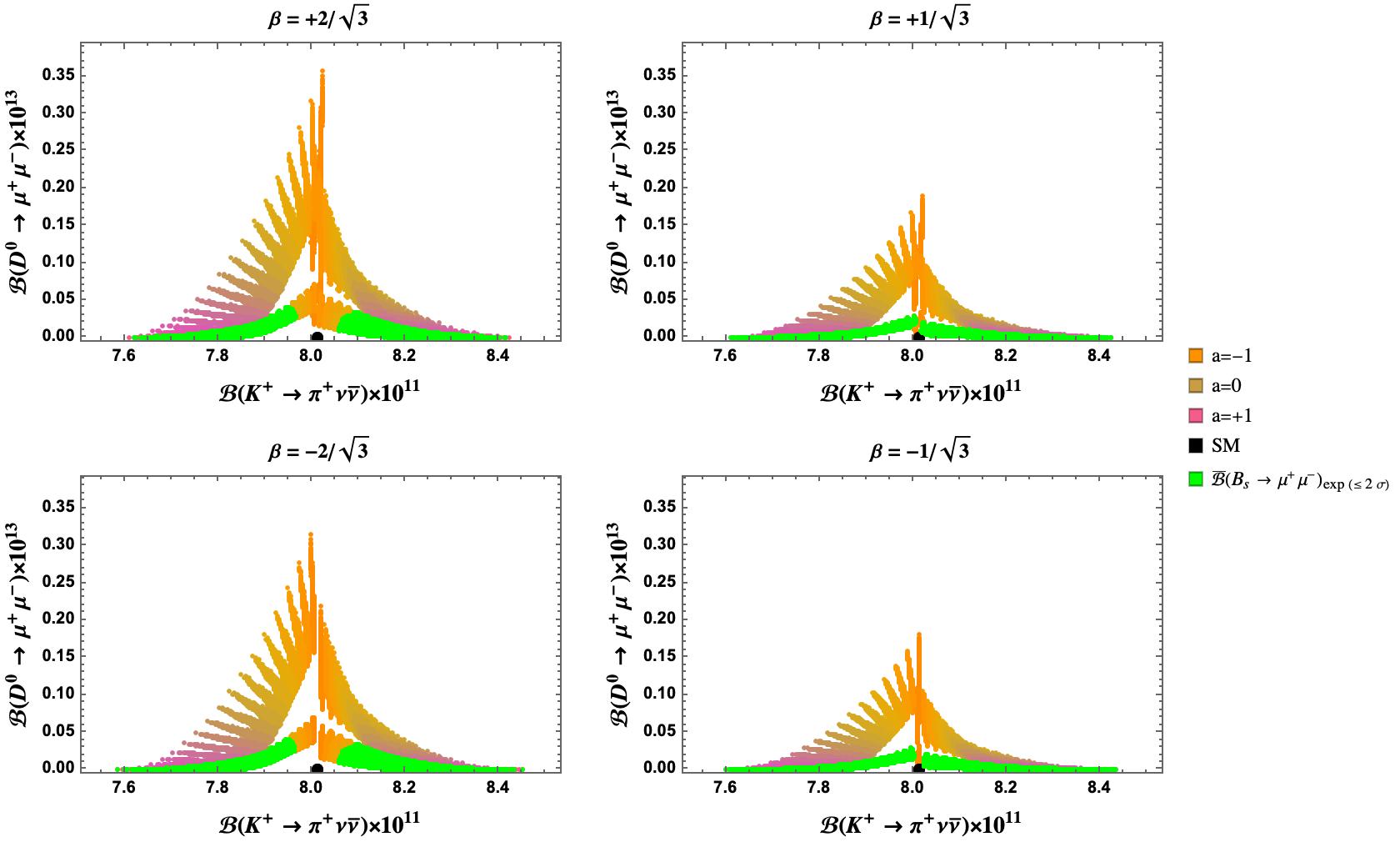}
    \caption{\small Correlation between ${\cal B}(D^0 \to \mu^+ \mu^-)$ and ${\cal B}(K^+ \to \pi^+  \nu \bar \nu)$ in 331 models with parameters as in Fig.\ref{plD0xD}.
 The black dot represents the SM result. The green points are obtained requiring that $\overline {\cal B}(B_s \to \mu^+ \mu^-)$ lies in the experimental range within $2\sigma$.  }\label{plkpiuD0mumu}
\end{center}
\end{figure}
\begin{figure}[!tb]
\begin{center}
\includegraphics[width = 0.9\textwidth]{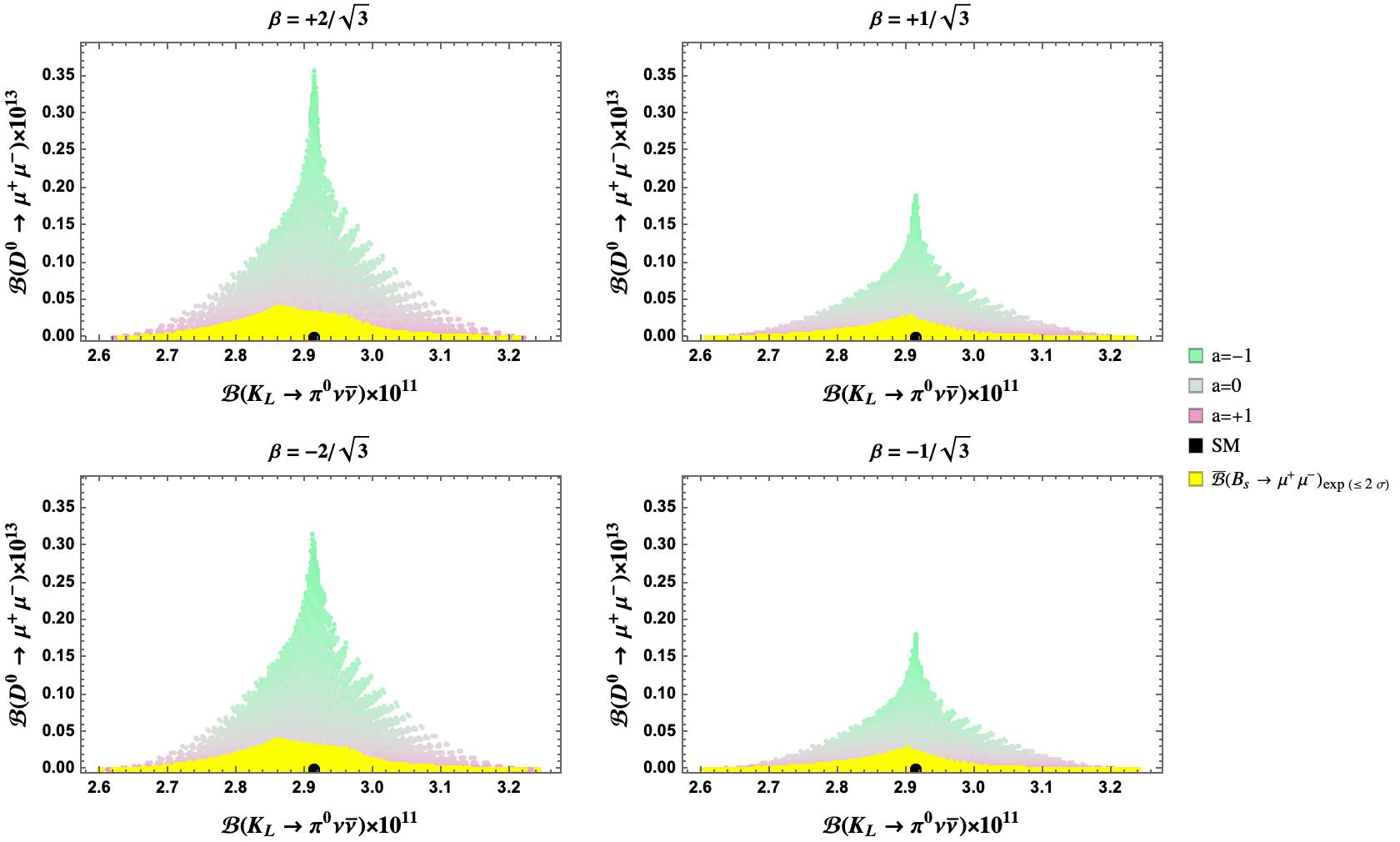}
    \caption{\small Correlation between ${\cal B}(D^0 \to \mu^+ \mu^-)$ and ${\cal B}(K_L \to \pi^0  \nu \bar \nu)$ in 331 models with parameters as in Fig.~\ref{plD0xD}.
The black dot represents the SM result. The yellow points are obtained requiring $\overline {\cal B}(B_s \to \mu^+ \mu^-)$ in the experimental range within $2\sigma$. }\label{plk0D0mumu}
\end{center}
\end{figure}

\begin{figure}[!tb]
\begin{center}
\includegraphics[width = 0.9\textwidth]{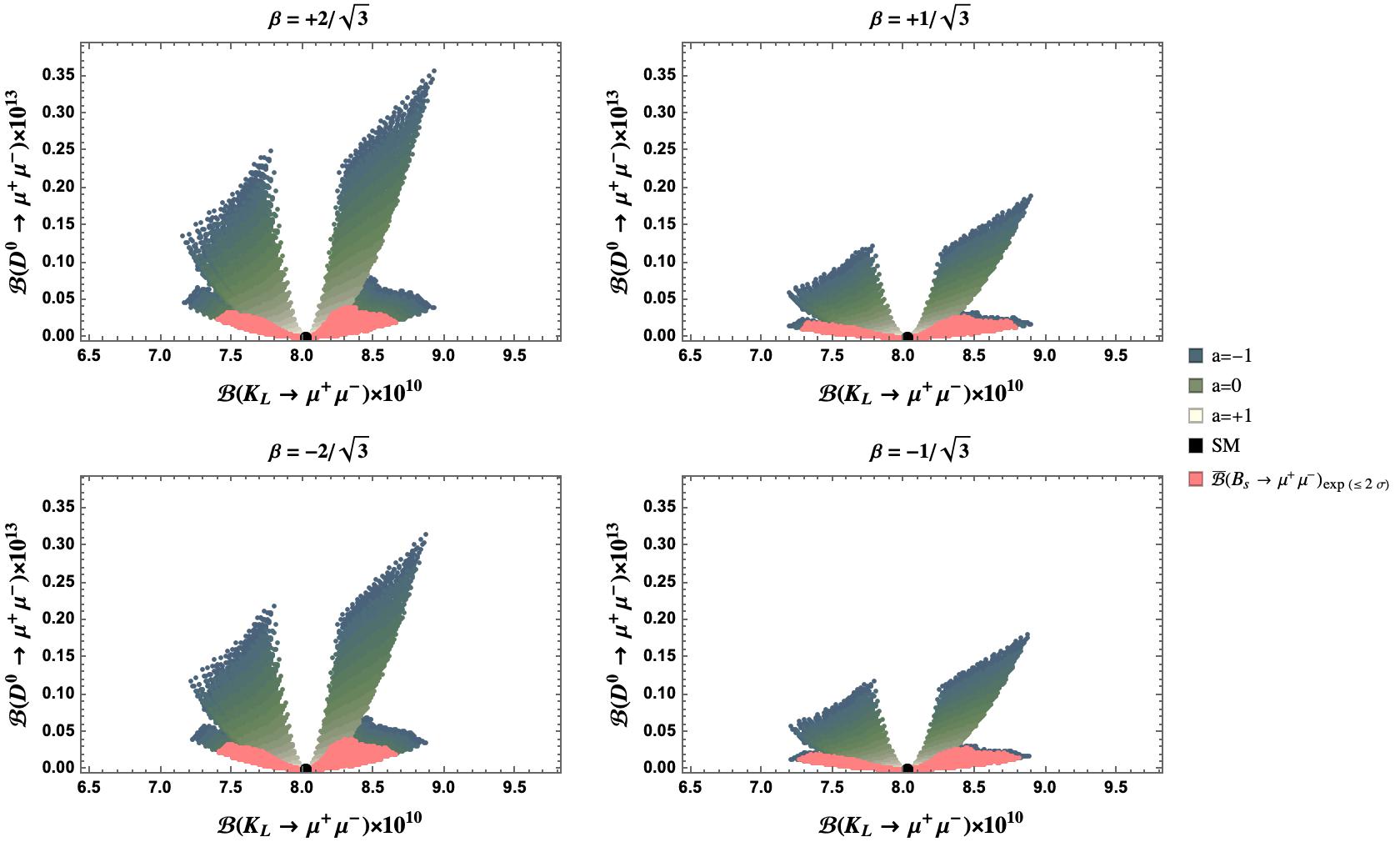}
    \caption{\small Correlation between ${\cal B}(D^0 \to \mu^+ \mu^-)$ and ${\cal B}(K_L \to \mu^+ \mu^-)$ in 331 models with parameters as in Fig.~\ref{plD0xD}.
 The black dot represents the SM result. The pink points are obtained requiring  $\overline {\cal B}(B_s \to \mu^+ \mu^-)$  in the experimental range within $2\sigma$. }\label{plkLmumuD0mumu}
\end{center}
\end{figure}
\begin{figure}[tb]
\begin{center}
\includegraphics[width = 0.9\textwidth]{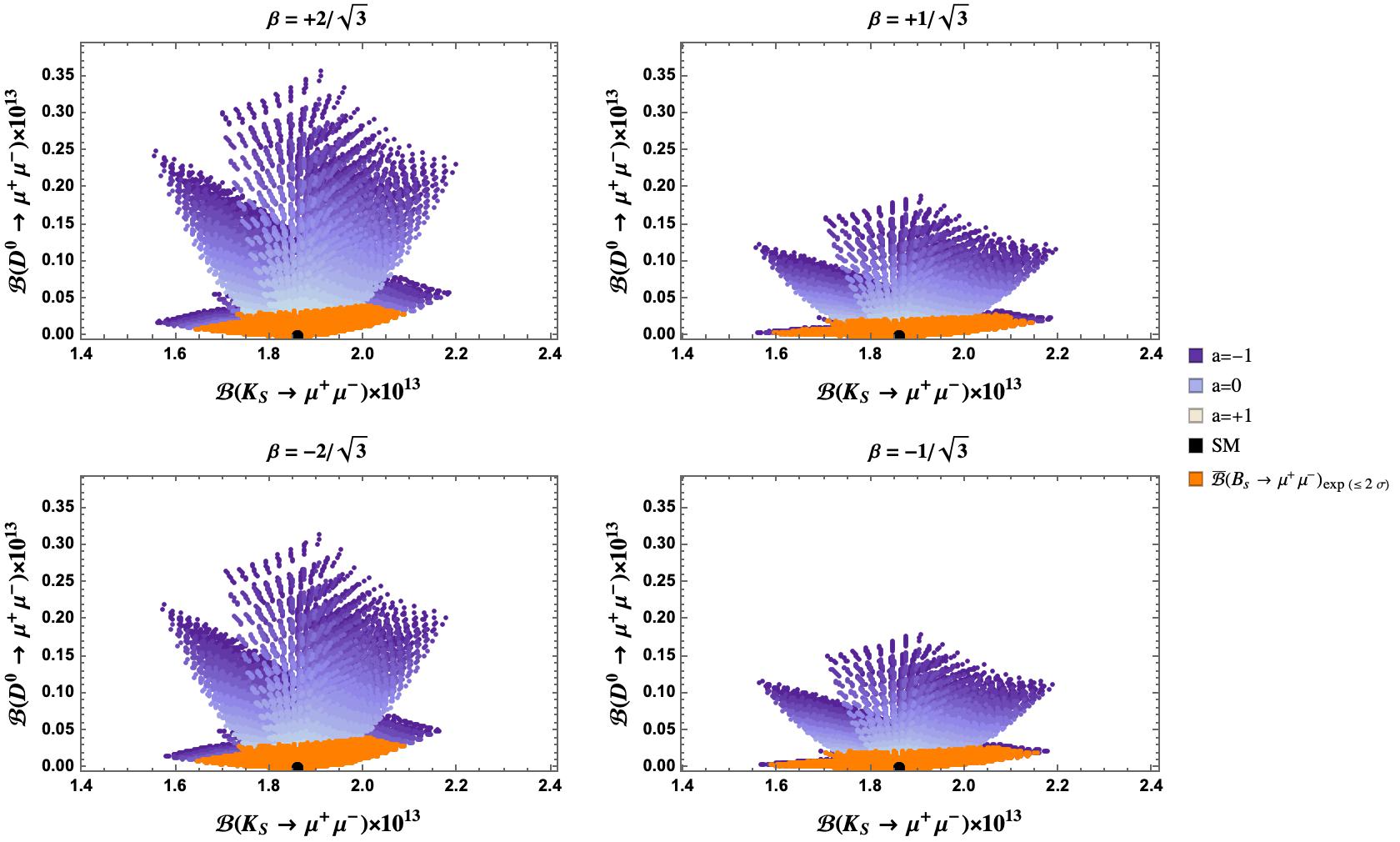}
    \caption{\small Correlation between ${\cal B}(D^0 \to \mu^+ \mu^-)$ and ${\cal B}(K_S \to \mu^+ \mu^-)$  in 331 models with parameters as in Fig.~\ref{plD0xD}.
The black dot represents the SM result. The orange points are obtained requiring  $\overline {\cal B}(B_s \to \mu^+ \mu^-)$ in the experimental range within $2\sigma$.  }\label{plkSmumuD0mumu}
\end{center}
\end{figure}

\appendix
\section{$\Delta F=1$ effective Hamiltonian and RG evolution of the coefficients\label{Operators}}
The most general effective Hamiltonian governing the decays  $D \to K^+ K^-$, $\pi^+ \pi^-$ can be written as
\bea
{\cal H}_{eff}^{\Delta F=1}&=&\sum_D \lambda_c^D \sum_{i=1}^2 \left( C_i^D Q_i^D+  {\tilde C}_i^D {\tilde Q}_i^D \right) +\sum_{j }\left( C_j Q_j+  {\tilde C}_j {\tilde Q}_j \right) \nn \\
&+&\sum_A \sum_D  \left( C_A^D Q_A^D+  {\tilde C}_A^D{\tilde Q}_A^D \right)+{\rm h.c.}
\eea
where $D=d,\,s$, $\lambda_c^D=V_{cD}^* V_{uD}$, $j \in [3,\dots 10,\, 8g]$ and  $A\in[S1,\,S2,\,T1,\,T2]$.
The operators are classified according to  the standard nomenclature:
\begin{itemize}
\item
current-current operators:
\be
Q_1^D=({\bar u}D)_{V-A}({\bar D}c)_{V-A} \hskip 1cm Q_2^D=({\bar u}_\alpha D_\beta )_{V-A}({\bar D}_\beta c_\alpha)_{V-A} \label{cc}
\ee
\item{QCD penguins}
\bea
Q_3&=&({\bar u} c)_{V-A} \sum_q ({\bar q}q)_{V-A} \hskip 1cm Q_4=({\bar u}_\alpha c_\beta)_{V-A} \sum_q ({\bar q}_\beta q_\alpha)_{V-A}  \label{QCD34} \\
Q_5&=&({\bar u} c)_{V-A} \sum_q ({\bar q}q)_{V+A} \hskip 1cm Q_6=({\bar u}_\alpha c_\beta)_{V-A} \sum_q ({\bar q}_\beta q_\alpha)_{V+A}  \label{QCD56} 
\eea
\item{QED penguins}
\bea
Q_7&=&\frac{3}{2} ({\bar u} c)_{V-A} \sum_q e_q ({\bar q}q)_{V+A} \hskip 1cm Q_8=\frac{3}{2} ({\bar u}_\alpha c_\beta)_{V-A} \sum_q e_q({\bar q}_\beta q_\alpha)_{V+A}  \label{QED78} \\
Q_9&=&\frac{3}{2} ({\bar u} c)_{V-A} \sum_q e_q ({\bar q}q)_{V-A} \hskip 1cm Q_{10}=\frac{3}{2} ({\bar u}_\alpha c_\beta)_{V-A} \sum_q e_q({\bar q}_\beta q_\alpha)_{V-A}  \label{QED910}
\eea
\item{chromomagnetic operator}
\be
Q_{8g}=\frac{g_s}{8\pi^2} m_c {\bar u}_\alpha  \sigma^{\mu \nu}(1+\gamma_5) c_\beta T^a_{\alpha \beta} G^a_{\mu \nu} \label{8g}
\ee
\item{scalar and tensor operators}
\bea
Q_{S1}&=&({\bar u} P_L D )({\bar D}P_L c) \hskip 2.2cm Q_{S2}=({\bar u}_\alpha P_L D_\beta )({\bar D}_\beta P_L c_\alpha) \label{scalars}
\\
Q_{T1}&=&({\bar u} \sigma_{\mu \nu}P_L D )({\bar D}  \sigma^{\mu \nu}  P_L c) \hskip 1cm Q_{T2}=({\bar u}_\alpha \sigma_{\mu \nu} P_L D_\beta )({\bar D}_\beta \sigma^{\mu \nu} P_L c_\alpha) \, .  \label{tensors}
\eea
\end{itemize}
The operators ${\tilde Q}_i$ are obtained exchanging $L \leftrightarrow R$.
In the SM  only $Q_1^D$ is  present at tree level; the operators $Q_2^D$, the QCD penguins and $O_{8g}$ are generated at ${\cal O}(\alpha_s)$; the QED penguins are negligible because of their small coefficients,  and the
scalar and tensor operators are absent.
In 331 models, at the scale $M_{Z^\prime}$  the tree-level $Z^\prime$ exchange produces the operators $Q_3$ and $Q_7$.
When evolved at the low scale $\mu_c \simeq m_c$ other operators are generated.
Neglecting the scalar and tensor operators (absent in SM and in 331 model), the RG evolution can be done separately for the 10  operators in Eqs.~(\ref{cc}), (\ref{QCD34})-(\ref{QCD56}), (\ref{QED78})-(\ref{QED910}) and for $Q_{8g}$.
The anomalous dimension matrix (ADM) for the set of 10 operators is:
%
\begin{equation}
{\hat \gamma}^0_{1,...,10} = \begin{pmatrix}  
\frac{-6}{N_c} & 6 & \frac{-2}{3N_c} & \frac{2}{3} & \frac{-2}{3N_c} &
\frac{2}{3} & 0 & 0 & 0 & 0 \\
6 & \frac{-6}{N_c} & 0 & 0 & 0 & 0 & 0 & 0 & 0 & 0 \\
0 & 0 & \frac{-22}{3N_c} & \frac{22}{3} & \frac{-4}{3N_c} &
\frac{4}{3} & 0 & 0 & 0 & 0 \\
0 & 0 & 6 - \frac{2N_f}{3N_c} & \frac{-6}{N_c}+\frac{2N_f}{3} &
\frac{-2N_f}{3N_c} & \frac{2N_f}{3} & 0 & 0 & 0 & 0 \\
0 & 0 & 0 & 0 & \frac{6}{N_c} & -6 & 0 & 0 & 0 & 0 \\
0 & 0 & \frac{-2N_f}{3N_c} & \frac{2N_f}{3} & \frac{-2N_f}{3N_c} &
\frac{6(1-N_c^2)}{N_c} + \frac{2N_f}{3} & 0 & 0 & 0 & 0 \\
0 & 0 & 0 & 0 & 0 & 0 & \frac{6}{N_c} & -6 & 0 & 0 \\
0 & 0 & \frac{N_d-2N_u}{3N_c} & \frac{2N_u-N_d}{3} & \frac{N_d-2N_u}{3N_c} &
\frac{2N_u-N_d}{3} & 0 & \frac{6(1-N_c^2)}{N_c} & 0 & 0 \\
0 & 0 & \frac{2}{3N_c} & -\frac{2}{3} & \frac{2}{3N_c} & -\frac{2}{3}
& 0 & 0 & \frac{-6}{N_c} & 6 \\
0 & 0 & \frac{N_d-2N_u}{3N_c} & \frac{2N_u-N_d}{3} & \frac{N_d-2N_u}{3N_c} &
\frac{2N_u-N_d}{3} & 0 & 0 & 6 & \frac{-6}{N_c}
\end{pmatrix} ~,
\end{equation}
%
where $N_c =3$ is the number of colors, $N_f$, $N_u$ and $N_d$ are the number of active
quark flavors, active  up- and active
down-type  quarks, respectively.
For the chromomagnetic operator $Q_{8g}^{(1)}$, the LO anomalous dimension reads 
\begin{equation}
{\hat \gamma}^0_{8g} = \frac{4 N_c^2 -8}{N_c} ~.
\end{equation}
The result of the evolution of the 331 operators leads to the generation of the operators having the structure of $Q_3-Q_8$, with coefficients given by:
\bea
C_3(\mu_c) &=&1.456 \, C_3(M_{Z^\prime})+0.012 \, C_7(M_{Z^\prime})  \label{C3-331} \\
C_4(\mu_c) &=&-0.7909 \, C_3(M_{Z^\prime})-0.026 \, C_7(M_{Z^\prime})  \label{C4-331} \\
C_5(\mu_c) &=&0.0085 \, C_3(M_{Z^\prime})+0.0074 \, C_7(M_{Z^\prime})  \label{C5-331} \\
C_6(\mu_c) &=&-0.1107 \, C_3(M_{Z^\prime})-0.0349 \, C_7(M_{Z^\prime})  \label{C6-331} \\
C_7(\mu_c) &=&R \, C_7(M_{Z^\prime}) \simeq 0.8244\, C_7(M_{Z^\prime})\label{C7-331} \\
C_8(\mu_c) &=&\frac{1- R^9 }{3 \,R^8}\,C_7(M_{Z^\prime}) \simeq 1.2879 \, C_7(M_{Z^\prime}) \, , \label{C8-331}
\eea
where $R=r(\mu_t,\,M_{Z^\prime})^{1/7}\, r(\mu_b,\,\mu_t)^{3/23}\, r(\mu_c,\,\mu_b)^{3/25}$ and $r(\mu_1,\,\mu_2)=\displaystyle\frac{\alpha_s(\mu_2)}{\alpha_s(\mu_1)}$. 
 $C_3(M_{Z^\prime})$ and $C_7(M_{Z^\prime})$ are those in Eqs.~(\ref{C3Zp})-(\ref{C7Zp}).
\bibliographystyle{JHEP}
\bibliography{refAFFP}

\providecommand{\href}[2]{#2}\begingroup\raggedright\begin{thebibliography}{10}

\bibitem{Wilkinson:2021tby}
G.~Wilkinson, {\it {Charming synergies: the role of charm-threshold studies in
  the search for physics beyond the Standard Model}},
  \href{http://arxiv.org/abs/2107.08414}{{\tt arXiv:2107.08414}}.

\bibitem{Lenz:2020awd}
A.~Lenz and G.~Wilkinson, {\it {Mixing and $CP$ violation in the charm
  system}},  \href{http://arxiv.org/abs/2011.04443}{{\tt arXiv:2011.04443}}.

\bibitem{Altmannshofer:2021qrr}
W.~Altmannshofer and P.~Stangl, {\it {New Physics in Rare B Decays after
  Moriond 2021}},  \href{http://arxiv.org/abs/2103.13370}{{\tt
  arXiv:2103.13370}}.

\bibitem{Alguero:2021anc}
M.~Alguer\'o, B.~Capdevila, S.~Descotes-Genon, J.~Matias, and M.~Novoa-Brunet,
  {\it {$\boldsymbol{b\to s\ell\ell}$ global fits after Moriond 2021 results}},
   in {\em {55th Rencontres de Moriond on QCD and High Energy Interactions}},
  4, 2021.
\newblock \href{http://arxiv.org/abs/2104.08921}{{\tt arXiv:2104.08921}}.

\bibitem{Colangelo:2021myn}
P.~Colangelo, F.~De~Fazio, and F.~Loparco, {\it {$c \to u \nu {\bar \nu}$
  transitions of $B_c$ mesons: 331 model facing Standard Model null tests}},
  \href{http://arxiv.org/abs/2107.07291}{{\tt arXiv:2107.07291}}.

\bibitem{Pisano:1991ee}
F.~Pisano and V.~Pleitez, {\it {An $SU(3) \times U(1)$ model for electroweak
  interactions}},  {\em Phys. Rev. D} {\bf 46} (1992) 410--417,
  [\href{http://arxiv.org/abs/hep-ph/9206242}{{\tt hep-ph/9206242}}].

\bibitem{Frampton:1992wt}
P.~H. Frampton, {\it {Chiral dilepton model and the flavor question}},  {\em
  Phys. Rev. Lett.} {\bf 69} (1992) 2889--2891.

\bibitem{Buras:2012dp}
A.~J. Buras, F.~De~Fazio, J.~Girrbach, and M.~V. Carlucci, {\it {The Anatomy of
  Quark Flavour Observables in 331 Models in the Flavour Precision Era}},  {\em
  JHEP} {\bf 02} (2013) 023, [\href{http://arxiv.org/abs/1211.1237}{{\tt
  arXiv:1211.1237}}].

\bibitem{Buras:2013dea}
A.~J. Buras, F.~De~Fazio, and J.~Girrbach, {\it {331 models facing new $b \to
  s\mu^+ \mu^-$ data}},  {\em JHEP} {\bf 02} (2014) 112,
  [\href{http://arxiv.org/abs/1311.6729}{{\tt arXiv:1311.6729}}].

\bibitem{Buras:2014yna}
A.~J. Buras, F.~De~Fazio, and J.~Girrbach-Noe, {\it {$Z$-$Z'$ mixing and
  $Z$-mediated FCNCs in $SU(3)_{C} \times SU(3)_{L} \times U(1)_{X}$ models}},
  {\em JHEP} {\bf 08} (2014) 039, [\href{http://arxiv.org/abs/1405.3850}{{\tt
  arXiv:1405.3850}}].

\bibitem{Buras:2015kwd}
A.~J. Buras and F.~De~Fazio, {\it {$\varepsilon'/\varepsilon$ in 331 Models}},
  {\em JHEP} {\bf 03} (2016) 010, [\href{http://arxiv.org/abs/1512.02869}{{\tt
  arXiv:1512.02869}}].

\bibitem{Buras:2016dxz}
A.~J. Buras and F.~De~Fazio, {\it {331 Models Facing the Tensions in $\Delta
  F=2$ Processes with the Impact on $\varepsilon^\prime/\varepsilon$,
  $B_s\to\mu^+\mu^-$ and $B\to K^*\mu^+\mu^-$}},  {\em JHEP} {\bf 08} (2016)
  115, [\href{http://arxiv.org/abs/1604.02344}{{\tt arXiv:1604.02344}}].

\bibitem{Cabarcas:2009vb}
J.~M. Cabarcas, D.~Gomez~Dumm, and R.~Martinez, {\it {Flavor changing neutral
  currents in 331 models}},  {\em J. Phys. G} {\bf 37} (2010) 045001,
  [\href{http://arxiv.org/abs/0910.5700}{{\tt arXiv:0910.5700}}].

\bibitem{Machado:2013jca}
A.~C.~B. Machado, J.~C. Montero, and V.~Pleitez, {\it {Flavor-changing neutral
  currents in the minimal 3-3-1 model revisited}},  {\em Phys. Rev. D} {\bf 88}
  (2013), no.~11 113002, [\href{http://arxiv.org/abs/1305.1921}{{\tt
  arXiv:1305.1921}}].

\bibitem{Cogollo:2012ek}
D.~Cogollo, A.~V. de~Andrade, F.~S. Queiroz, and P.~Rebello~Teles, {\it {Novel
  sources of Flavor Changed Neutral Currents in the $331_{RHN}$ model}},  {\em
  Eur. Phys. J. C} {\bf 72} (2012) 2029,
  [\href{http://arxiv.org/abs/1201.1268}{{\tt arXiv:1201.1268}}].

\bibitem{Amhis:2019ckw}
{\bf HFLAV} Collaboration, Y.~S. Amhis et~al., {\it {Averages of b-hadron,
  c-hadron, and $\tau $-lepton properties as of 2018}},  {\em Eur. Phys. J. C}
  {\bf 81} (2021), no.~3 226, [\href{http://arxiv.org/abs/1909.12524}{{\tt
  arXiv:1909.12524}}].

\bibitem{Aaij:2021aam}
{\bf LHCb} Collaboration, R.~Aaij et~al., {\it {Observation of the mass
  difference between neutral charm-meson eigenstates}},
  \href{http://arxiv.org/abs/2106.03744}{{\tt arXiv:2106.03744}}.

\bibitem{Bhardwaj:2019vep}
V.~Bhardwaj, M.~Dorigo, and F.-S. Yu, {\it {Summary of WG7 at CKM 2018: "Mixing
  and $CP$ violation in the $D$ system: $x_D$, $y_D$, $|q/p|_D$, $\phi_D$, and
  direct $CP$ violation in $D$ decays"}},  2019.
\newblock \href{http://arxiv.org/abs/1901.08131}{{\tt arXiv:1901.08131}}.

\bibitem{Cerri:2018ypt}
A.~Cerri, V.~V. Gligorov, S.~Malvezzi, J.~Martin~Camalich, and J.~Zupan, {\it
  {Opportunities in Flavour Physics at the HL-LHC and HE-LHC}},
  \href{http://arxiv.org/abs/1812.07638}{{\tt arXiv:1812.07638}}.

\bibitem{Gabbiani:1996hi}
F.~Gabbiani, E.~Gabrielli, A.~Masiero, and L.~Silvestrini, {\it {A Complete
  analysis of FCNC and CP constraints in general SUSY extensions of the
  standard model}},  {\em Nucl. Phys.} {\bf B477} (1996) 321--352,
  [\href{http://arxiv.org/abs/hep-ph/9604387}{{\tt hep-ph/9604387}}].

\bibitem{Bona:2007vi}
{\bf UTfit} Collaboration, M.~Bona et~al., {\it {Model-independent constraints
  on $\Delta F=2$ operators and the scale of new physics}},  {\em JHEP} {\bf
  0803} (2008) 049, [\href{http://arxiv.org/abs/0707.0636}{{\tt
  arXiv:0707.0636}}]. {Updates available on \texttt{http://www.utfit.org}.}

\bibitem{Isidori:2010kg}
G.~Isidori, Y.~Nir, and G.~Perez, {\it {Flavor Physics Constraints for Physics
  Beyond the Standard Model}},  {\em Ann.Rev.Nucl.Part.Sci.} {\bf 60} (2010)
  355, [\href{http://arxiv.org/abs/1002.0900}{{\tt arXiv:1002.0900}}].

\bibitem{Charles:2013aka}
J.~Charles, S.~Descotes-Genon, Z.~Ligeti, S.~Monteil, M.~Papucci, et~al., {\it
  {Future sensitivity to new physics in $B_d$, $B_s$ and $K$ mixings}},  {\em
  Phys.~Rev.} {\bf D89} (2014) 033016,
  [\href{http://arxiv.org/abs/1309.2293}{{\tt arXiv:1309.2293}}].

\bibitem{Golowich:2007ka}
E.~Golowich, J.~Hewett, S.~Pakvasa, and A.~A. Petrov, {\it {Implications of
  $D^0$ - $\bar{D}^0$ Mixing for New Physics}},  {\em Phys. Rev. D} {\bf 76}
  (2007) 095009, [\href{http://arxiv.org/abs/0705.3650}{{\tt
  arXiv:0705.3650}}].

\bibitem{Golowich:2009ii}
E.~Golowich, J.~Hewett, S.~Pakvasa, and A.~A. Petrov, {\it {Relating $D^0 -
  \bar D^0$ Mixing and $D^0 \to l^+ l^-$ with New Physics}},  {\em Phys. Rev.
  D} {\bf 79} (2009) 114030, [\href{http://arxiv.org/abs/0903.2830}{{\tt
  arXiv:0903.2830}}].

\bibitem{Buras:2000if}
A.~J. Buras, M.~Misiak, and J.~Urban, {\it {Two loop QCD anomalous dimensions
  of flavor changing four quark operators within and beyond the standard
  model}},  {\em Nucl. Phys. B} {\bf 586} (2000) 397--426,
  [\href{http://arxiv.org/abs/hep-ph/0005183}{{\tt hep-ph/0005183}}].

\bibitem{Bigi:2009df}
I.~I. Bigi, M.~Blanke, A.~J. Buras, and S.~Recksiegel, {\it {CP Violation in
  $D^0 - {\bar D}^0$ Oscillations: General Considerations and Applications to
  the Littlest Higgs Model with T-Parity}},  {\em JHEP} {\bf 07} (2009) 097,
  [\href{http://arxiv.org/abs/0904.1545}{{\tt arXiv:0904.1545}}].

\bibitem{Grossman:2009mn}
Y.~Grossman, Y.~Nir, and G.~Perez, {\it {Testing New Indirect CP Violation}},
  {\em Phys. Rev. Lett.} {\bf 103} (2009) 071602,
  [\href{http://arxiv.org/abs/0904.0305}{{\tt arXiv:0904.0305}}].

\bibitem{Aaij:2019kcg}
{\bf LHCb} Collaboration, R.~Aaij et~al., {\it {Observation of CP Violation in
  Charm Decays}},  {\em Phys. Rev. Lett.} {\bf 122} (2019), no.~21 211803,
  [\href{http://arxiv.org/abs/1903.08726}{{\tt arXiv:1903.08726}}].

\bibitem{Golden:1989qx}
M.~Golden and B.~Grinstein, {\it {Enhanced CP Violations in Hadronic Charm
  Decays}},  {\em Phys. Lett. B} {\bf 222} (1989) 501--506.

\bibitem{Grossman:2006jg}
Y.~Grossman, A.~L. Kagan, and Y.~Nir, {\it {New physics and CP violation in
  singly Cabibbo suppressed D decays}},  {\em Phys. Rev. D} {\bf 75} (2007)
  036008, [\href{http://arxiv.org/abs/hep-ph/0609178}{{\tt hep-ph/0609178}}].

\bibitem{Brod:2011re}
J.~Brod, A.~L. Kagan, and J.~Zupan, {\it {Size of direct CP violation in singly
  Cabibbo-suppressed D decays}},  {\em Phys. Rev. D} {\bf 86} (2012) 014023,
  [\href{http://arxiv.org/abs/1111.5000}{{\tt arXiv:1111.5000}}].

\bibitem{Pirtskhalava:2011va}
D.~Pirtskhalava and P.~Uttayarat, {\it {CP Violation and Flavor SU(3) Breaking
  in D-meson Decays}},  {\em Phys. Lett. B} {\bf 712} (2012) 81--86,
  [\href{http://arxiv.org/abs/1112.5451}{{\tt arXiv:1112.5451}}].

\bibitem{Bhattacharya:2012ah}
B.~Bhattacharya, M.~Gronau, and J.~L. Rosner, {\it {CP asymmetries in
  singly-Cabibbo-suppressed $D$ decays to two pseudoscalar mesons}},  {\em
  Phys. Rev. D} {\bf 85} (2012) 054014,
  [\href{http://arxiv.org/abs/1201.2351}{{\tt arXiv:1201.2351}}].

\bibitem{Feldmann:2012js}
T.~Feldmann, S.~Nandi, and A.~Soni, {\it {Repercussions of Flavour Symmetry
  Breaking on CP Violation in D-Meson Decays}},  {\em JHEP} {\bf 06} (2012)
  007, [\href{http://arxiv.org/abs/1202.3795}{{\tt arXiv:1202.3795}}].

\bibitem{Cheng:2012wr}
H.-Y. Cheng and C.-W. Chiang, {\it {Direct CP violation in two-body hadronic
  charmed meson decays}},  {\em Phys. Rev. D} {\bf 85} (2012) 034036,
  [\href{http://arxiv.org/abs/1201.0785}{{\tt arXiv:1201.0785}}]. [Erratum:
  Phys.Rev.D 85, 079903 (2012)].

\bibitem{Franco:2012ck}
E.~Franco, S.~Mishima, and L.~Silvestrini, {\it {The Standard Model confronts
  CP violation in $D^0 \to \pi^+\pi^-$ and $D^0 \to K^+K^-$}},  {\em JHEP} {\bf
  05} (2012) 140, [\href{http://arxiv.org/abs/1203.3131}{{\tt
  arXiv:1203.3131}}].

\bibitem{Cheng:2012xb}
H.-Y. Cheng and C.-W. Chiang, {\it {$SU(3)$ symmetry breaking and CP violation
  in $D \to PP$ decays}},  {\em Phys. Rev. D} {\bf 86} (2012) 014014,
  [\href{http://arxiv.org/abs/1205.0580}{{\tt arXiv:1205.0580}}].

\bibitem{Cheng:2019ggx}
H.-Y. Cheng and C.-W. Chiang, {\it {Revisiting CP violation in $D\to P\!P$ and
  $V\!P$ decays}},  {\em Phys. Rev. D} {\bf 100} (2019), no.~9 093002,
  [\href{http://arxiv.org/abs/1909.03063}{{\tt arXiv:1909.03063}}].

\bibitem{Brod:2012ud}
J.~Brod, Y.~Grossman, A.~L. Kagan, and J.~Zupan, {\it {A Consistent Picture for
  Large Penguins in $D\to\pi^+\pi^-,K^+K^-$}},  {\em JHEP} {\bf 10} (2012) 161,
  [\href{http://arxiv.org/abs/1203.6659}{{\tt arXiv:1203.6659}}].

\bibitem{Isidori:2011qw}
G.~Isidori, J.~F. Kamenik, Z.~Ligeti, and G.~Perez, {\it {Implications of the
  LHCb Evidence for Charm CP Violation}},  {\em Phys. Lett. B} {\bf 711} (2012)
  46--51, [\href{http://arxiv.org/abs/1111.4987}{{\tt arXiv:1111.4987}}].

\bibitem{Giudice:2012qq}
G.~F. Giudice, G.~Isidori, and P.~Paradisi, {\it {Direct CP violation in charm
  and flavor mixing beyond the SM}},  {\em JHEP} {\bf 04} (2012) 060,
  [\href{http://arxiv.org/abs/1201.6204}{{\tt arXiv:1201.6204}}].

\bibitem{Altmannshofer:2012ur}
W.~Altmannshofer, R.~Primulando, C.-T. Yu, and F.~Yu, {\it {New Physics Models
  of Direct CP Violation in Charm Decays}},  {\em JHEP} {\bf 04} (2012) 049,
  [\href{http://arxiv.org/abs/1202.2866}{{\tt arXiv:1202.2866}}].

\bibitem{Hiller:2012wf}
G.~Hiller, Y.~Hochberg, and Y.~Nir, {\it {Supersymmetric $\Delta A_{CP}$}},
  {\em Phys. Rev. D} {\bf 85} (2012) 116008,
  [\href{http://arxiv.org/abs/1204.1046}{{\tt arXiv:1204.1046}}].

\bibitem{Chala:2019fdb}
M.~Chala, A.~Lenz, A.~V. Rusov, and J.~Scholtz, {\it {$\Delta A_{CP}$ within
  the Standard Model and beyond}},  {\em JHEP} {\bf 07} (2019) 161,
  [\href{http://arxiv.org/abs/1903.10490}{{\tt arXiv:1903.10490}}].

\bibitem{Li:2019hho}
H.-N. Li, C.-D. Lü, and F.-S. Yu, {\it {Implications on the first observation
  of charm CPV at LHCb}},  \href{http://arxiv.org/abs/1903.10638}{{\tt
  arXiv:1903.10638}}.

\bibitem{Grossman:2019xcj}
Y.~Grossman and S.~Schacht, {\it {The emergence of the $\Delta U=0$ rule in
  charm physics}},  {\em JHEP} {\bf 07} (2019) 020,
  [\href{http://arxiv.org/abs/1903.10952}{{\tt arXiv:1903.10952}}].

\bibitem{Dery:2019ysp}
A.~Dery and Y.~Nir, {\it {Implications of the LHCb discovery of CP violation in
  charm decays}},  {\em JHEP} {\bf 12} (2019) 104,
  [\href{http://arxiv.org/abs/1909.11242}{{\tt arXiv:1909.11242}}].

\bibitem{Khodjamirian:2017zdu}
A.~Khodjamirian and A.~A. Petrov, {\it {Direct CP asymmetry in $D\to
  \pi^-\pi^+$ and $D\to K^-K^+$ in QCD-based approach}},  {\em Phys. Lett. B}
  {\bf 774} (2017) 235--242, [\href{http://arxiv.org/abs/1706.07780}{{\tt
  arXiv:1706.07780}}].

\bibitem{Fajfer:2015mia}
S.~Fajfer and N.~Ko\v{s}nik, {\it {Prospects of discovering new physics in rare
  charm decays}},  {\em Eur. Phys. J. C} {\bf 75} (2015), no.~12 567,
  [\href{http://arxiv.org/abs/1510.00965}{{\tt arXiv:1510.00965}}].

\bibitem{Buras:2020xsm}
A.~Buras, {\em {Gauge Theory of Weak Decays}}.
\newblock Cambridge University Press, 6, 2020.

\bibitem{Zyla:2020zbs}
{\bf Particle Data Group} Collaboration, P.~Zyla et~al., {\it {Review of
  Particle Physics}},  {\em PTEP} {\bf 2020} (2020), no.~8 083C01.

\bibitem{Burdman:2001tf}
G.~Burdman, E.~Golowich, J.~L. Hewett, and S.~Pakvasa, {\it {Rare charm decays
  in the standard model and beyond}},  {\em Phys. Rev. D} {\bf 66} (2002)
  014009, [\href{http://arxiv.org/abs/hep-ph/0112235}{{\tt hep-ph/0112235}}].

\bibitem{Aoki:2019cca}
{\bf Flavour Lattice Averaging Group} Collaboration, S.~Aoki et~al., {\it {FLAG
  Review 2019: Flavour Lattice Averaging Group (FLAG)}},  {\em Eur. Phys. J. C}
  {\bf 80} (2020), no.~2 113, [\href{http://arxiv.org/abs/1902.08191}{{\tt
  arXiv:1902.08191}}].

\bibitem{Chetyrkin:2017lif}
K.~G. Chetyrkin, J.~H. Kuhn, A.~Maier, P.~Maierhofer, P.~Marquard,
  M.~Steinhauser, and C.~Sturm, {\it {Addendum to \textquotedblleft{}Charm and
  bottom quark masses: An update\textquotedblright{}}},
  \href{http://arxiv.org/abs/1710.04249}{{\tt arXiv:1710.04249}}. [Addendum:
  Phys.Rev.D 96, 116007 (2017)].

\bibitem{Chetyrkin:2009fv}
K.~G. Chetyrkin, J.~H. Kuhn, A.~Maier, P.~Maierhofer, P.~Marquard,
  M.~Steinhauser, and C.~Sturm, {\it {Charm and Bottom Quark Masses: An
  Update}},  {\em Phys. Rev. D} {\bf 80} (2009) 074010,
  [\href{http://arxiv.org/abs/0907.2110}{{\tt arXiv:0907.2110}}].

\bibitem{Buras:1990fn}
A.~J. Buras, M.~Jamin, and P.~H. Weisz, {\it {Leading and Next-to-leading {QCD}
  Corrections to $\epsilon$ Parameter and $B^0 - \bar{B}^0$ Mixing in the
  Presence of a Heavy Top Quark}},  {\em Nucl. Phys. B} {\bf 347} (1990)
  491--536.

\bibitem{Urban:1997gw}
J.~Urban, F.~Krauss, U.~Jentschura, and G.~Soff, {\it {Next-to-leading order
  QCD corrections for the $B^0 - \bar B^0$ mixing with an extended Higgs
  sector}},  {\em Nucl. Phys. B} {\bf 523} (1998) 40--58,
  [\href{http://arxiv.org/abs/hep-ph/9710245}{{\tt hep-ph/9710245}}].

\bibitem{Gupta:1996yt}
R.~Gupta, T.~Bhattacharya, and S.~R. Sharpe, {\it {Matrix elements of four
  fermion operators with quenched Wilson fermions}},  {\em Phys. Rev. D} {\bf
  55} (1997) 4036--4054, [\href{http://arxiv.org/abs/hep-lat/9611023}{{\tt
  hep-lat/9611023}}].

\bibitem{Lin:2006vc}
H.-W. Lin, S.~Ohta, A.~Soni, and N.~Yamada, {\it {Charm as a domain wall
  fermion in quenched lattice QCD}},  {\em Phys. Rev. D} {\bf 74} (2006)
  114506, [\href{http://arxiv.org/abs/hep-lat/0607035}{{\tt hep-lat/0607035}}].

\bibitem{DeBruyn:2012wk}
K.~De~Bruyn, R.~Fleischer, R.~Knegjens, P.~Koppenburg, M.~Merk, A.~Pellegrino,
  and N.~Tuning, {\it {Probing New Physics via the $B^0_s\to \mu^+\mu^-$
  Effective Lifetime}},  {\em Phys. Rev. Lett.} {\bf 109} (2012) 041801,
  [\href{http://arxiv.org/abs/1204.1737}{{\tt arXiv:1204.1737}}].

\end{thebibliography}\endgroup
\end{document}